\documentclass[aps,twocolumn,a4paper,floatfix]{revtex4}
\usepackage{graphicx}
\usepackage{subfigure}
\usepackage{adjustbox}
\usepackage{bm}
\usepackage{color}
\usepackage{braket}
\usepackage{standalone}
\usepackage{multirow}
\usepackage{tikz}
\usepackage{mathrsfs}
\usepackage{dsfont}
\usepackage[colorlinks,bookmarks=true,citecolor=blue,linkcolor=red,urlcolor=blue]{hyperref}
\usepackage{amsmath}
\usepackage{amssymb}

\begin{document}
\title{Sensitivity of non-Hermitian systems}

\author{Elisabet Edvardsson}
\affiliation{Department of Physics, Stockholm University, AlbaNova University Center, 106 91 Stockholm, Sweden}
\author{Eddy Ardonne}
\affiliation{Department of Physics, Stockholm University, AlbaNova University Center, 106 91 Stockholm, Sweden}

\date{\today}

\begin{abstract}
Understanding the extreme sensitivity of the eigenvalues of non-Hermitian Hamiltonians to the boundary conditions is of great importance when analyzing non-Hermitian systems, as it appears generically and is intimately connected to the skin effect and the  breakdown of the conventional bulk boundary correspondence. Here we describe a method to find the eigenvalues of one-dimensional one-band models with arbitrary boundary conditions. We use this method on several systems to find analytical expressions for the eigenvalues, which give us conditions on the parameter values in the system for when we can expect the spectrum to be insensitive to a change in  boundary conditions. By stacking one-dimensional chains, we use the derived results to find corresponding conditions for insensitivity for some two-dimensional systems with periodic boundary conditions in one direction.
This would be hard by using other methods to detect skin effect, such as the winding of the determinant of the Bloch Hamiltonian. Finally, we use these results to make predictions about the (dis)appearance of the skin effect in purely two-dimensional systems with open boundary conditions in both directions.

\end{abstract}

\maketitle

\section{Introduction}

In the past few years, the interest in non-Hermitian Hamiltonians, and in particular the topological properties of such systems \cite{BeBuKu2019,OkKaShSa2020,KaBeSa2019,LeLiGo2019,EdKuBe2019,Yoshida2019,KoBu2020,GoAsKaTaHiUe2018,ZhLe2019,LeBlHuChNo2017,ShZhFu2018,KaShUeSa2019,YaWa2018,KuEdBuBe2018,AlVaTo2018,Xi2018,Le2016,KuDw2019,XiDeWaZhWaYiXu2019,DeYoHa2021,MaBe2021,StBe2021,VyRo2021,RiRo2022}, has become huge. The differences between non-Hermitian systems and their Hermitian counterparts are many and give rise to interesting effects.

Mathematically, these differences arise from the fact that a non-Hermitian operator can lose several of the key properties that make Hermitian operators so nice to work with. Most importantly, the eigenvalues are no longer necessarily real, the right and left eigenvectors might not be related to each other via Hermitian conjugation or be orthogonal amongst themselves. Also, the operator might not be diagonalizable, leading to the existence of exceptional points. This makes computing probabilities and expectation values non-trivial, something which is discussed in e.g. \cite{Br2014}, where the \emph{biorthogonal inner product} is defined and used to replace the standard inner product in probability calculations. 

In particular, there are several new phenomena that occur in non-Hermitian tight-binding models compared to Hermitian ones. Perhaps most notably, the bulk-boundary correspondence breaks down \cite{YaWa2018,EdKuBe2019,KuEdBuBe2018,AlVaTo2018,Xi2018,Le2016, KoBu2020,KuDw2019}, and is replaced by a biorthogonal bulk-boundary correspondence as described in \cite{KuEdBuBe2018,EdKuBe2019}. The breakdown of the bulk-boundary correspondence is related to the extreme sensitivity of eigenvalues and eigenvectors to boundary conditions that often occurs in non-Hermitian systems. In fact, in a system with open boundary conditions, the left and right eigenstates tend to be exponentially localized to one of the boundaries of the system, and the number of such localized states is extensive. This phenomenon is called the \emph{skin effect} \cite{YaWa2018}. The breakdown of the bulk-boundary correspondence and the skin effect are intimately connected to each other and have lead to an active field of research \cite{HeBaRe2019,Lo2019,LeTh2019,LiZhAiGoKaUeNo2019,FlZoVaBeBaTi2022,ZiReRo2021,BoKrSl2020,YoMu2019,YaMoBe2021,EdKuYoBe2020,Sc2020,BrHy2019}. In particular, the phenomena have been experimentally verified in mechanical systems \cite{GhBrWeCo2019,BrLoLeCo2019,BrViRoAr2021,RoRu2020}, topoelectrical circuits \cite{HeHoImAbKiMoLeSzGrTh2019,HoHeScSaBrGrKiWoVoKaLeBiThNe2020}, photonic systems \cite{XiDeWaZhWaYiXu2019,WeKrHeHoStGrThSz2020}, and in ultracold atoms \cite{LiXiDoLiLiGaYiYa2022}. This phenomenology has also been suggested to be of practical use in sensors whose sensitivity increases exponentially with the size of the system \cite{BuBe2020,KoBu2022,DoCl2020}. 

In one-dimensional systems it is clear what is meant by this description of the skin effect, but in higher dimensions a little more care is needed in defining what is meant as states can be localized to different boundaries of the system in different ways.  Nevertheless, the fact that most of the eigenstates are exponentially localized to some boundary of the open system, gives an intuitive explanation of why the spectrum and the eigenstates should change radically when coupling the ends -- if the boundary is removed, the states cannot be localized to it anymore and the exponential localization must disappear. Because the eigenstates drastically change when coupling the edges, the eigenvalues also undergo a large shift.

Now, it turns out that not all non-Hermitian systems exhibit a skin effect, and one important question is how to determine which systems do and which not. In a one-dimensional system, described by the Bloch Hamiltonian $H(k)$, it turns out that there is a relation between the winding number 

\begin{equation}
\label{eq:winding}
w(E_B) = \frac{1}{2\pi i}\int_{-\pi}^{\pi}\frac{d}{dk}\ln\det[H(k)-E_B] dk,
\end{equation}
around some base energy $E_B\in \mathbb{C}$ and the existence of the skin effect in the corresponding system with open boundary conditions. Namely, if there exists an $E_B\in \mathbb{C}$ such that $w(E_B)\neq 0$, then the system with open boundary conditions has a skin effect \cite{GoAsKaTaHiUe2018,OkKaShSa2020}, otherwise not. This also implies that if we have a non-zero winding number for some $E_B$, the spectrum of the system will be very sensitive to boundary conditions. The winding number is related to the types of gaps that can be found in a non-Hermitian system \cite{GoAsKaTaHiUe2018,KaShUeSa2019}. Namely, if $w(E_B)$ is non-zero for some $E_B$, we say that the system exhibits a point gap. This shows up as loops in the spectrum of the Bloch Hamiltonian. On the other hand, if $w(E_B) = 0$ for all $E_B$, we say we have a line gap, which typically implies that the spectrum of the Bloch Hamiltonian consists of line segments that do not form loops.  

In \cite{KuEdBuBe2018}, it was numerically found that if the ends of an SSH-chain are coupled by a parameter $\delta$, the $\delta$ required to change the spectrum by a fixed amount $\Delta$ is proportional to $e^{-\xi(\Delta) N}$, where $N$ is the length of the chain and $\xi(\Delta)>0$ also depends on system parameters.
This kind of sensitivity was also studied in \cite{KoBu2020}, and we will call it \emph{exponential sensitivity} to boundary conditions, because an exponentially small change in the boundary condition $\delta$ leads to a finite change $\Delta$ in the spectrum.
If we do not have this kind of sensitivity, we say the spectrum has \emph{non-exponential sensitivity}.
Clearly, when numerically calculating the spectrum of non-Hermitian operators, one should be careful if the system has exponential sensitivity to the boundary conditions. For larger system sizes, even a machine-precision deviation can lead to substantial errors in the eigenvalues obtained. It is thus advantageous to know in advance when the system is exponentially sensitive and when it is not. We note that algorithms calculating the spectrum of non-Hermitian PT-symmetric systems to arbitrary precision we considered in \cite{NoLuJe2013,NoLuStJe2017}.

In this paper, we aim to study the sensitivity of the eigenvalues analytically for different systems described by tight-binding Hamiltonians of the form
\begin{equation}
    \mathcal{H} = \sum_{mn} t_{mn}c_{m}^{\dagger}c_n, 
\end{equation}
where, $c_{m}^{\dagger}$ and $c_{n}$ are creation and annihilation operators respectively and the $t_{mn}$ are hopping parameters for which $t_{mn}$ is not necessarily equal to $t_{nm}^*$. We begin with purely one-dimensional tight-binding models where we interpolate between open and periodic boundary conditions using a parameter $\delta$. For a class of such systems, we develop a method to analytically find how the eigenvalues depend on the parameter $\delta$ and the system size. Then we move on to two-dimensional systems, constructed by stacking one-dimensional chains. In a rectangular geometry, the Hamiltonian can be represented by a block-tridiagonal block-Toeplitz matrix, with extra blocks in the corners to account for the boundary conditions. Such matrices can in general not be diagonalized exactly, unless we have periodic boundary conditions, but using the result from the one-dimensional case, we can still draw some conclusions about these systems.

From the analytical expressions for the eigenvalues, we can see for which parameter values the spectrum should be exponentially sensitive and for which not. It turns out that in order for the spectrum to have non-exponential sensitivity there has to be some kind of balancing of the hopping parameters; somewhat loosely we need the amount of hopping to the right to be balanced by a similar amount to the left. Because of this, we will say that a system with parameter values such that the spectrum has non-exponential sensitivity is \emph{balanced}. We find that the parameter values for which the system is balanced are in agreement with what the winding number predicts about the skin effect. For a model, with the hopping parameters explicitly specified, one can often easily obtain the winding number by looking at the plot of $H(k)$ in the complex plane. However, finding out for which parameters the model has (or does not have) a non-trivial winding, is often much more complicated. In those cases, it might be much easier to determine if the system is balanced or not.

The paper is organized as follows. In Sec.~\ref{sec:one-dimensioal-systems}, we introduce the method to analyze the non-Hermitian systems we are interested in, and apply it to several models, including the Hatano-Nelson model and the SSH-chain. In Sec.~\ref{sec:two-dimensional-systems}, we use solved one-dimensional systems, and construct several two-dimensional models, including the triangular lattice, which shows particularly interesting behavior. We discuss the results in Sec.~\ref{sec:discussion}. The appendices provide more details on some of the models we studied.

\section{One-dimensional systems}
\label{sec:one-dimensioal-systems}

We begin by describing a method to find the eigenvalues of a one-dimensional, one-band, non-Hermitian system.
Any such system with $N$ sites and a maximum hopping range of $m<\lfloor N/2 \rfloor$, can be described by an $N\times N$ Toeplitz matrix. Now, in order to simplify the notation, we will here only show the method for a system with next-nearest neighbour hopping, but it is straightforwardly generalizable to include longer range hopping. The Hamiltonian of a one-dimensional system with next-nearest neighbour hopping and the coupling strength between the ends determined by the parameter $\delta$ can be represented by the following matrix: 

\begin{equation}
	H = \begin{pmatrix}
	t_{11}       & t_{12}       & t_{13} &        & \delta t_{31}       &\delta t_{21}\\
	t_{21}       & \ddots       & \ddots & \ddots &                     &\delta t_{31}\\
	t_{31}       & \ddots       & \ddots & \ddots & \ddots              &             \\
	             & \ddots       & \ddots & \ddots & \ddots              & t_{13}\\
	\delta t_{13}&              & \ddots & \ddots & \ddots              & t_{12}\\
	\delta t_{12}& \delta t_{13}&        & t_{31} & t_{21}              & t_{11}	
	\end{pmatrix} \ .
\end{equation}

For $\delta=1$, we have a system with periodic boundary conditions and for $\delta= 0$, we have a system with open boundary conditions. In the case of periodic boundary conditions, the eigenvalues are found by Fourier transforming the system, and are given by
\begin{equation}
	\lambda_k^{\delta=1} = t_{11}+t_{12}\omega^{k}+t_{13}\omega^{2k}+t_{21}\omega^{-k}+t_{31}\omega^{-2k},
\end{equation}
where $k=0,1,\dots,N-1$ and $\omega= \exp(2\pi i/N)$. Now, we make the ansatz that the eigenvalues for the case with general $\delta$ are given by
\begin{equation}\label{eq:eigenvalue_ansatz}
\lambda_{\alpha} =  t_{11}+t_{12}e^{i\alpha}+t_{13}e^{2i\alpha}+t_{21}e^{-i\alpha}+t_{31}e^{-2i\alpha},
\end{equation}
where $\alpha$ is a complex number, the possible values of which need to be determined.

From the Schrödinger equation, $H\psi_{\alpha} = \lambda_{\alpha} \psi_{\alpha}$, we get a system of equations containing $N-4$ \emph{bulk equations} of the form
\begin{align}\label{eq:bulk_eqs}
&	t_{31}\psi_{\alpha,n-2}+t_{21}\psi_{\alpha,n-1} +(t_{11}-\lambda_{\alpha})\psi_{\alpha,n} \nonumber\\
&    +t_{12}\psi_{\alpha,n+1}+t_{13}\psi_{\alpha,n+2} = 0
\end{align}
with $n = 3,\dots,N-2$, and four \emph{boundary equations} 
\begin{equation}\label{eq:boundary_eqs}
	\begin{cases}
	\!\begin{aligned}
	(t_{11}-\lambda_{\alpha})\psi_{\alpha,1}+t_{12}\psi_{\alpha,2}&+t_{13}\psi_{\alpha,3}+\delta t_{31}\psi_{\alpha,N-1}\\&+\delta t_{21}\psi_{\alpha,N} = 0,
	\end{aligned}\\
	\!\begin{aligned}
	t_{21}\psi_{\alpha,1}+(t_{11}-\lambda_{\alpha})\psi_{\alpha,2}&+t_{12}\psi_{\alpha,3}+t_{13}\psi_{\alpha,4}\\&+\delta t_{31}\psi_{\alpha,N} = 0,
	\end{aligned}\\
	\!\begin{aligned}
	\delta t_{13}\psi_{\alpha,1}+t_{31}\psi_{\alpha,N-3}&+t_{21}\psi_{\alpha,N-2}+(t_{11}-\lambda_{\alpha})\psi_{\alpha,N-1}\\&+t_{12}\psi_{\alpha,N} = 0,
	\end{aligned}\\
	\!\begin{aligned}
	\delta t_{12}\psi_{\alpha,1}+\delta t_{13}\psi_{\alpha,2}&+t_{31}\psi_{\alpha,N-2}+t_{21}\psi_{\alpha,N-1}\\&+(t_{11}-\lambda_{\alpha})\psi_{\alpha,N}=0 \ .
	\end{aligned}
	\end{cases}
\end{equation}
The strategy to find the eigenvalues is to first find the general solution of Eq.~\eqref{eq:bulk_eqs}
for {\em arbitrary} integer $n$, and then to determine for which values of $\alpha$ there exists a non-trivial
solution to the boundary equations in Eq.~\eqref{eq:boundary_eqs}.
Because we solve the bulk equation in Eq.~\eqref{eq:bulk_eqs} for arbitrary integer $n$, we can use it to simplify the
boundary equations, which gives
\begin{equation}
\begin{split}
\label{eq:alt_boundary_eqs}
t_{31} (\delta\psi_{\alpha,N-1}-\psi_{\alpha,-1}) + t_{21} (\delta\psi_{\alpha,N}-\psi_{\alpha,0}) & = 0, \\
t_{31} (\delta\psi_{\alpha,N}-\psi_{\alpha,0}) & = 0, \\
t_{13} (\delta\psi_{\alpha,1}-\psi_{\alpha,N+1}) & = 0, \\
t_{12} (\delta\psi_{\alpha,1}-\psi_{\alpha,N+1}) + t_{13} (\delta\psi_{\alpha,2}-\psi_{\alpha,N+2}) & = 0.
\end{split}
\end{equation}

Eq.~\eqref{eq:bulk_eqs} is a linear recurrence relation for the elements of the vector $\psi_{\alpha}$ and has a characteristic polynomial of the form
\begin{equation}
	x^4+\frac{t_{12}}{t_{13}}x^3+\frac{t_{11}-\lambda_{\alpha}}{t_{13}}x^2+\frac{t_{21}}{t_{13}}x+\frac{t_{31}}{t_{13}} = 0,
\end{equation}
for $t_{13}\neq 0$. (In case $t_{13} = 0$, we get a polynomial of lower degree.) The zeros of this polynomial, $x_i$, give us the general solution of the recurrence relation, namely
\begin{equation}
	\psi_{\alpha,n} = \sum_i c_i x_i^n,
\end{equation} 
where the constants $c_i$ are determined by inserting the expression for $\psi_{\alpha,n}$ into the boundary equations in Eq.~\eqref{eq:alt_boundary_eqs}. This gives us a linear system of equations for the constants of the form
\begin{equation}
	M(\alpha)C = 0,
\end{equation} 
where $M(\alpha)$ is a $4\times 4$-matrix and $C$ is the vector containing the constants $c_i$. To find the values of $\alpha$ for which this system has a non-trivial solution, we compute the determinant of the matrix $M(\alpha)$ and solve the equation
\begin{equation}
	\det M(\alpha) =0.
\end{equation}
The form of this equation can tell us how sensitive $\alpha$, and in turn the eigenvalues $\lambda_{\alpha}$ are to changes in the boundary conditions.

In general, the bulk and boundary equations cannot be solved analytically as this involves finding roots of polynomials of high degree, but for some systems the method provides useful information.
We point out that similar methods were used in, for instance, \cite{GuLiZhLiCh2021,Lo2019a} in the special cases of the Hatano-Nelson model and the SSH-chain. In these papers, an explicit ansatz was used for the eigenvectors, instead of using the recurrence relation.

To analyze the results, we are interested in the properties of the eigenvectors.
For non-Hermitian systems, the left and right eigenvectors are generically not orthonormal, and one can therefore consider different ways in which to normalize them. The eigenvectors considered above, $\psi_\alpha$, with components $\psi_{\alpha,n}$, are the right eigenvectors, and we denote those by
$\psi^r_\alpha $ and $\psi^r_{\alpha,n}$, respectively, if confusion with the left eigenvectors can arise.
The left eigenvectors, $\psi^l_\alpha$ with components $\psi^l_{\alpha,n}$, are obtained by diagonalizing $H^T$ instead of $H$.
In the cases we consider below, this means a simple swap of parameters of the model.

Following \cite{Br2014,KuEdBuBe2018,EdKuBe2019}, we consider different expectation values of the site projection operator $\Pi_n = c^\dagger_n c_n$.
On the one hand we have the left and right expectation values
$\langle \Pi_{n} \rangle_\alpha^{l,l} = (\psi_{\alpha}^l)^{\dagger} \Pi_{n}  \psi_{\alpha}^l  = |\psi_{\alpha,n}^l |^2$
and  
$\langle \Pi_{n} \rangle_\alpha^{r,r} = (\psi_{\alpha}^r)^{\dagger} \Pi_{n} \psi_{\alpha}^r = |\psi_{\alpha,n}^r|^2$,
which can be used to determine if the system exhibits skin effect. Namely, the right and left expectation values tell us where in the system the eigenvectors are localized, which, as explained in the introduction tells us if the skin effect is present or not. In the case of one-dimensional systems, as we consider in this section, skin effect simply means that an extensive number of right eigenvectors are localized on one side of the system, while the left eigenvectors are localized near the other.

The third kind of expectation value we consider, is the biorthogonal expectation value, which, as is shown in \cite{Br2014} is the one most similar to what we call an expectation value in Hermitian quantum mechanics. The biorthogonal expectation value is given by
$\langle \Pi_{n} \rangle_\alpha^{l,r} = (\psi_{\alpha}^l)^{\dagger} \Pi_n  \psi_{\alpha}^r = (\psi_{\alpha,n}^l)^* \psi_{\alpha,n}^r$
and can be used to distinguish between bulk and boundary states in the system \cite{KuEdBuBe2018,EdKuBe2019}.

\subsection{The Hatano-Nelson model}
\label{sec:hn-model}
We begin with a simple, non-Hermitian nearest-neighbour hopping model, also known as the Hatano-Nelson model \cite{HaNe1996,HaNe1997,HaNe1998}.
The Hamiltonian for a chain of length $N$ with boundary conditions regulated by the parameter $\delta$ is given by

\begin{align}
	\mathcal{H} &= \sum_{n= 1}^{N-1}
	\left[t_rc_{n+1}^{\dagger}c_n+t_lc_{n}^{\dagger}c_{n+1}+t_d c_n^{\dagger}c_n\right] + \\\nonumber
	&+t_d c_N^{\dagger}c_N + \delta t_r c_{1}^{\dagger}c_N+\delta t_l c_{N}^{\dagger}c_{1} \ .
\end{align}

To analyze the situation where we interpolate between open and periodic boundary conditions, it is easier to write the Hamiltonian in matrix form. Then, it is given by the $N\times N$ matrix
\begin{equation}
	H = \begin{pmatrix}
	t_d& t_l & & \delta t_r\\
	t_r &\ddots &\ddots &\\
	& \ddots &\ddots & t_l\\
	\delta t_l & & t_r & t_d
	\end{pmatrix},
	\label{eq:very_simple_hamiltonian}
\end{equation}
where $\delta = 1$ corresponds to periodic boundary conditions and $\delta = 0$ corresponds to open boundary conditions. Below, we only give the Hamiltonians in matrix form as in Eq.~\eqref{eq:very_simple_hamiltonian}. Often, we consider the case $0 \leq \delta \leq 1$, but unless otherwise stated, the formulas are valid for arbitrary (complex) $\delta$. In Appendix~\ref{app:hn-more-general}, we derive the results for more general boundary conditions, where we allow for different coupling parameters in different directions. 

In accordance with Eq.~\eqref{eq:eigenvalue_ansatz}, we get the following ansatz for the eigenvalues in the case with arbitrary boundary conditions:
\begin{equation}
	\lambda_{\alpha} = t_d + t_l e^{i\alpha}+t_r e^{-i\alpha},
\end{equation}
where $\alpha$ is some complex number. The recurrence relation for the elements in the eigenvector is now given by
\begin{equation}
	t_r \psi_{\alpha,n-1} + t_l \psi_{\alpha,n+1} - (t_l e^{i \alpha} + t_r e^{-i \alpha}) \psi_{\alpha,n} = 0, \label{eq:recursive_HN}\\
\end{equation}
which has the solution
\begin{equation}
\psi_{\alpha,n} = c_1e^{in\alpha}+c_2r^n e^{-in\alpha},
\end{equation}
where $r = t_r/t_l$.
We see here that as long as $c_2\neq 0$ and $|r| \neq 1$, we have an exponential localization of all the eigenstates to one of the ends of the chain, depending on the absolute values of $t_r$ and $t_l$.
To find out when it is possible to have $c_2 = 0$ (that is, when we do not have an exponential localization of the eigenstates near the edges), we use the boundary equations, which in this case are given by
\begin{align}
\label{eq:HN-boundary}
t_r (\delta \psi_{\alpha,N} - \psi_{\alpha,0}) & = 0, &
t_l (\delta \psi_{\alpha,1} - \psi_{\alpha,N+1}) & = 0 \ .
\end{align}
By setting $c_2 = 0$, we can use Eq.~\eqref{eq:HN-boundary} to determine the values of $\delta$ for which we get $N$ inequivalent solutions for $\alpha$, giving a complete set of eigenvalues. This procedure leads to two possible values of $\delta$, namely $\delta = \pm 1$.
For other values of $\delta$, we have $c_2 \neq 0$.
Below, we give the explicit eigenstates for completeness, even though we do not need them to determine if the eigenstates are exponentially localized near the edges or not.

We continue by determining the equation for $\alpha$, such that we obtain a non-trivial solution for the coefficients $c_1$ and $c_2$. 
It turns out that it is convenient to write the eigenvalues in terms of 
\begin{equation}\label{eq:shifteq1}
	\tilde{\alpha} = \alpha+ i \ln\left( \frac{\sqrt{t_r}}{\sqrt{t_l}}\right) \ .
\end{equation}
The eigenvalues are then given by
\begin{equation}\label{eq:HNeigenvalues}
	\lambda_{\tilde\alpha} = t_d+2\sqrt{t_l}\sqrt{t_r}\cos(\tilde{\alpha}) \ ,
\end{equation}
while the eigenvectors read
\begin{equation}
\psi_{\tilde\alpha,n} = \Bigl(\frac{\sqrt{t_r}}{\sqrt{t_l}}\Bigr)^n \bigl(c_1e^{in\tilde\alpha}+c_2 e^{-in\tilde\alpha}\bigr) \ .
\end{equation}
In this form, one finds, provided the solutions for $\tilde\alpha$ are real, that the eigenvalues lie on a straight line in the complex plane.
Similar variable changes, though in slightly different settings, for example in the context of the generalized Brillouin zone, were considered in \cite{YaWa2018,KuDw2019}.
The possible values of $\tilde{\alpha}$ correspond to the solutions of the equation
\begin{equation}\label{eq:alphaeq}
\begin{split}
	&\delta\left(\frac{t_l^{N/2}}{t_r^{N/2}}+\frac{t_r^{N/2}}{t_l^{N/2}}\right)-\frac{\sin[(N+1)\tilde{\alpha}]}{\sin(\tilde{\alpha})}\\&+\delta^2\frac{\sin[(N-1)\tilde{\alpha}]}{\sin(\tilde{\alpha})} = 0 \ ,
\end{split}
\end{equation}
which is found by computing the determinant obtained from the boundary equations, as explaind above. We note that the form of the eigenvalues in Eq.\eqref{eq:HNeigenvalues} is similar, up to values of $\tilde{\alpha}$, to the eigenvalues of a tridiagonal Toeplitz matrix.

For completeness, we give a more explicit form of the (right) eigenvectors for arbitrary $\delta$, though still in terms of the parameter $\tilde\alpha$.
Solving the boundary Eq.~\eqref{eq:HN-boundary} results in
\begin{equation}
\psi_{\tilde\alpha,n} = \Bigl(\frac{\sqrt{t_r}}{\sqrt{t_l}}\Bigr)^n \Bigl(
\sin(n \tilde\alpha) + 
\delta \Bigl(\frac{\sqrt{t_r}}{\sqrt{t_l}}\Bigr)^N 
\sin[(N-n) \tilde\alpha]
\Bigr) \ .
\end{equation}
For $\delta = 0$, one can even obtain the possible values for $\tilde\alpha$ explicitly.
Eq.~\eqref{eq:alphaeq} for $\tilde\alpha$ then simplifies, with the (independent) solutions given by
$\tilde\alpha = \frac{\pi k'}{(N+1)}$, with $k' = 1,2,\ldots N$, resulting in
\begin{equation}
\label{eq:delta0}
\psi_{k',n} = \Bigl(\frac{\sqrt{t_r}}{\sqrt{t_l}}\Bigr)^n \sin\Bigl(\frac{n k' \pi}{N+1}\Bigr) \ .
\end{equation}
The left eigenvectors can be obtained by diagonalizing $H^T$, which means swapping $t_l \leftrightarrow t_r$.

We have now obtained expressions for the eigenvalues and eigenvectors, such that we can investigate the their behavior when we change $\delta$.

\subsubsection{Interpolating the boundary conditions}

\begin{figure}
	\includegraphics[width=\linewidth]{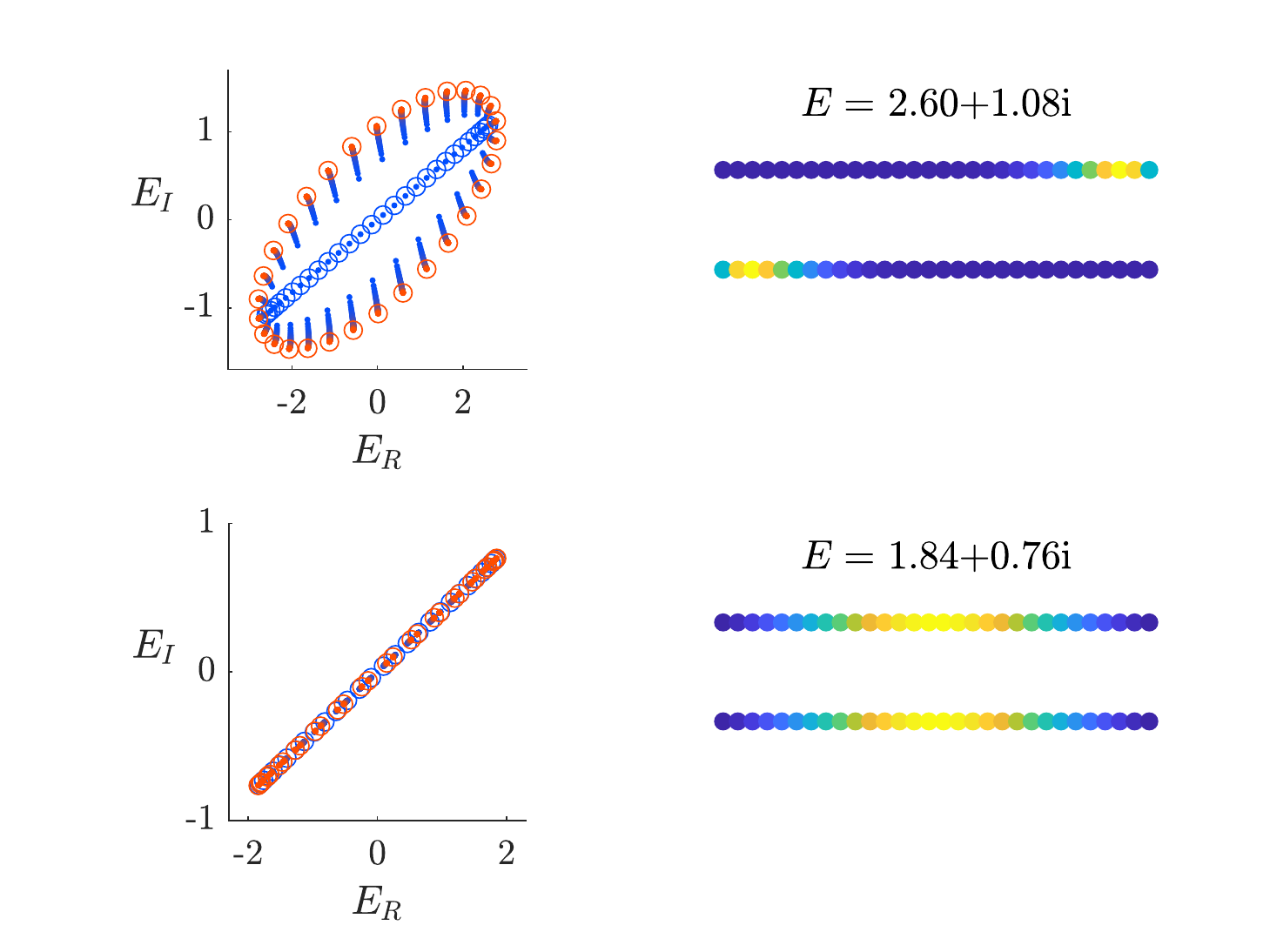}
	\caption{The left panel shows the eigenvalues of the Hatano-Nelson model with $N=30$ sites for $t_l = 1$ and $t_r = 2e^{i\pi/4}$ (upper left) and $t_l = 1$ and $t_r = e^{i\pi/4}$ (lower left). Red circles correspond to periodic boundary conditions and blue circles to open boundary conditions. The dots show how the eigenvalues change when we change $\delta$ in steps of $0.01$ from $0$ to $1$. The right panel shows the right and left expectation values of the projection operator in a representative state for both sets of parameters in the upper right and lower right respectively.
	}
	\label{fig:hatano_nelson_figure}
\end{figure}

We start by noting that as long as $|t_l|\neq |t_r|$, the equation for $\tilde{\alpha}$, Eq.~\eqref{eq:alphaeq}, contains a term proportional to $\delta$ that is exponential in the system size $N$. Therefore, the spectrum has an exponential sensitivity to the boundary conditions when $|t_r|\neq|t_l|$. 
When, on the other hand, $|t_l|=|t_r|$, the spectrum is non-exponentially sensitive to boundary conditions. We note that this happens when there seems to be an overall balancing of the hoppings in the system in the different directions, and therefore, as mentioned in the introduction, we will say that the hoppings are \emph{balanced} when this happens. In fact, when this is the case, $r$ is equal to a phase, say $e^{i\theta}$, which means that Eq.~\eqref{eq:alphaeq} can be written as
\begin{equation}
\label{eq:hn-alpha-tilde}
	2\delta\cos(\theta)\sin(\tilde{\alpha}) = \sin([N+1]\tilde{\alpha})-\delta^2\sin([N-1]\tilde{\alpha}).
\end{equation}

We show in Appendix~\ref{app:alpha_real} that for $\delta\in[-1,1]$ and $\theta$ real, this equation only has real solutions for $\tilde\alpha$. This implies that Eq.~\eqref{eq:HNeigenvalues} describes a line segment in the complex plane and that the system does not have a point gap at the balanced parameter values. Therefore, the system does not exhibit a skin effect when it is balanced, which is in correspondence with the fact that the spectrum has non-exponential sensitivity to boundary conditions at these parameter values.

We show this difference in behavior between balanced and unbalanced systems when going from periodic to open boundary conditions in Fig~\ref{fig:hatano_nelson_figure}.
In the upper left panel, we show the eigenvalues for the model with $N=30$ sites, and parameters
$t_l = 1$ and $t_r = 2 e^{i \pi/4}$ such that $|t_l| \neq |t_r|$.
In the lower left panel, we show the same results, but for the parameters $t_l = 1$ and $t_r = e^{i \pi/4}$ instead, such that $|t_l| = |t_r|$.
In the former case, we see a `jump' in the eigenvalues between $\delta = 0$ and $\delta = 1/100$, indicating a significant change in the spectrum, while in the latter case, such a jump is absent and the eigenvalues stay on the same line in the complex plane, as explained above.

In the right panel, we plot, for both sets of parameters, the left and right expectation values, $\langle \Pi_{\tilde\alpha,n} \rangle^{l,l} = | \psi^l_{\tilde\alpha,n}|^2$ and $\langle \Pi_{\tilde\alpha,n} \rangle^{r,r} = | \psi^r_{\tilde\alpha,n}|^2$, for a representative eigenstate,
showing a pronounced skin effect in the upper panel, while in the lower panel, the skin effect is absent.

\begin{figure}[t]
\includegraphics[width=\linewidth]{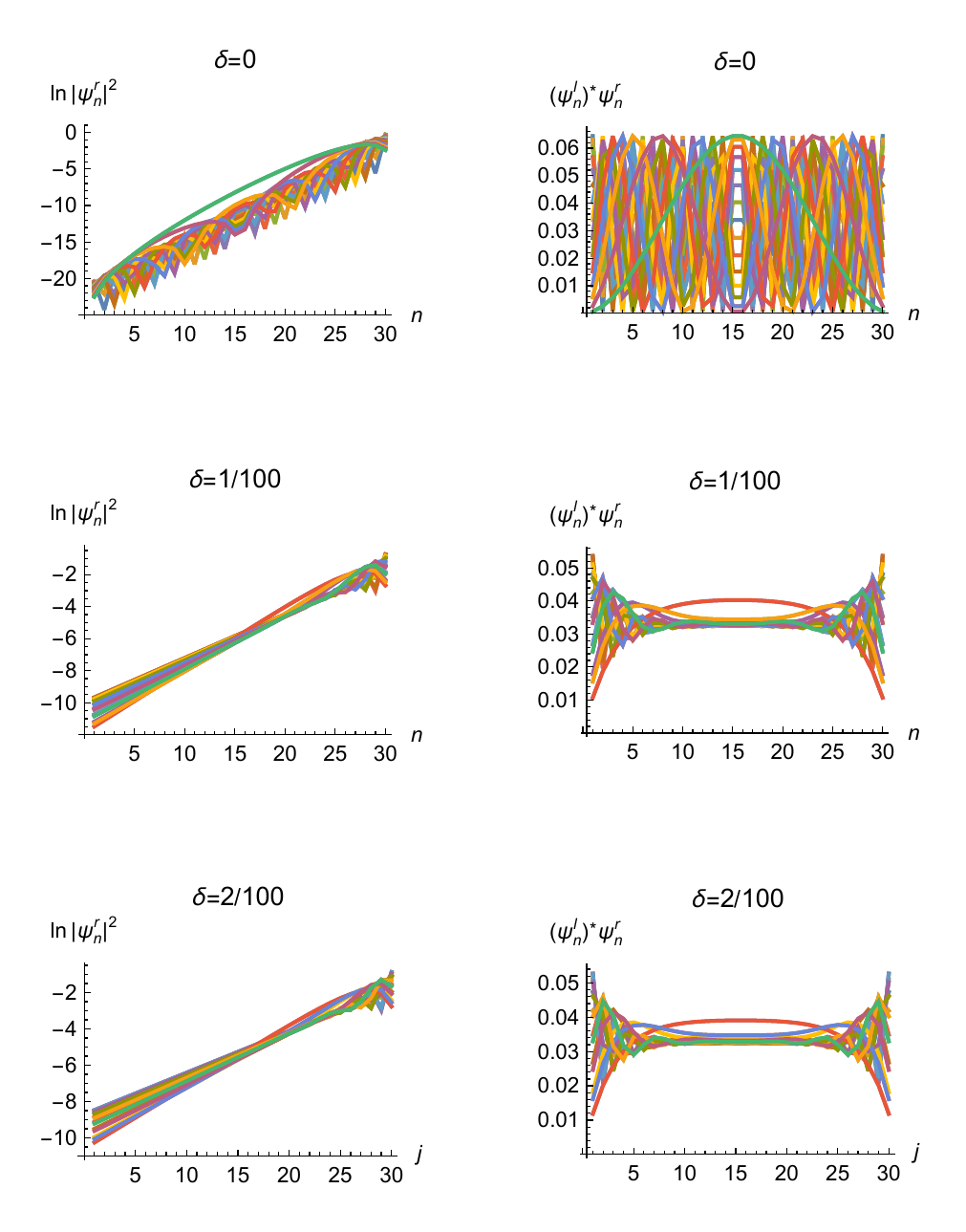}
\caption{Plot of the eigenvectors for the HN-model, with $N=30$ sites, $t_l = 1$, $t_r = 2e^{i \pi/4}$ 
and $\delta = 0$, $\delta = 1/100$ and $\delta = 2/100$ in the upper, middle and lower rows.
The left column shows the logarithm of the right expectation values, $\ln|\psi_n^r|^2$, while the right column shows the biorthogonal expectation values, $(\psi_{n}^l)^*\psi_n^r$.}
\label{fig:eigenvectors}
\end{figure}

The large change in the eigenvalues upon a small change in $\delta$ also shows up in the eigenvectors. 
In Fig.~\ref{fig:eigenvectors}, we plot both the (logarithm of the) right expectation values $\langle \Pi_{\tilde\alpha,n} \rangle^{r,r} = | \psi^r_{\tilde\alpha,n}|^2$, 
as well as all the biorthogonal expectation values $\langle \Pi_{\tilde\alpha,n} \rangle^{l,r} = (\psi^l_{\tilde\alpha,n})^* \psi^r_{\tilde\alpha,n}$.
The right eigenvectors clearly show the presence of the skin effect, because they are exponentially localised on the right hand side of the system.
For $\delta = 0$, there is a clear oscillatory behavior, which quickly disappears upon increasing $\delta$, as follows from Eq.~\eqref{eq:delta0}.

In addition, the correlation length changes by several orders of magnitude upon changing $\delta = 0$ to $\delta = 1/100$, while it remains almost the same upon changing $\delta = 1/100$ to $\delta = 2/100$.
The left eigenvectors show similar behavior, but are localised on the left hand side of the system.

The biorthogonal expectation value also shows the exponential sensitivity to small deviations of $\delta$ from zero.
In this case, the weight of the eigenvectors is spread out over the bulk of the system (so we do not have edge states), but the way in which this occurs changes drastically with $\delta$.
For $\delta = 0$, the inner product oscillates as a function of position (see Eq.~\eqref{eq:delta0}), while upon increasing $\delta$, these oscillations disappear in the bulk, where the biorthogonal inner product becomes constant.

In conclusion, we see a perfect correspondence between the exponential localisation of eigenstates to one side of the chain, the sensitivity of the eigenvalues to the boundary conditions and the existence of point gaps in the system.
In particular, we see that when $|t_r| = |t_l|$, we have a non-Hermitian system whose eigenvalues and eigenstates have properties that closely resemble what we have in a Hermitian system in terms of sensitivity to boundary conditions and localization respectively.

\subsubsection{Application to impurity-like systems}

We can also consider the case $\delta > 1$.
In this case, we do not think of $\delta$ as a parameter interpolating between open and periodic boundary conditions, but rather as a parameter that sets the strength of an impurity in the system, namely a single link with enhanced hopping.

For $\delta > 1$, we observe the following behavior:
Both for balanced $|t_l|=|t_r|$ as well as unbalanced parameters, $|t_l| \neq |t_r|$, there is one impurity state, signified by an exponential localization of the biorthogonal expectation around the impurity %
\footnote{Though the interpretation is very different, the mathematical description of this impurity state resembles the mathematical description of the topological zero mode that can be found in Hermitian systems.}.
The remaining states are bulk states.

\subsection{The SSH-chain}
\label{sec:ssh}

A system with more interesting properties than the Hatano-Nelson model, is the SSH-chain. It is described by the $N\times N$-matrix 
\begin{equation}
	H_0 = \begin{pmatrix}
	0& t_{l,1} & & &&\delta t_{r,2}\\
	t_{r,1} &0 &t_{l,2}& &&\\
	& t_{r,2} &\ddots &t_{l,1}&&\\
	 & & t_{r,1} & \ddots&\ddots&\\
	 &&&\ddots&\ddots&t_{l,1}\\
	 \delta t_{l,2}&&&&t_{r,1}&0
	\end{pmatrix},
	\label{eq:ssh_hamiltonian}
\end{equation}
where $\delta\in[0,1]$ and interpolates between open and periodic boundary conditions and $N$ determines the length of the chain. Here $N$ is assumed to be even since we want to be able to interpolate between open and periodic boundary conditions, but the case $N$ odd can be dealt with in a similar way, and the result is stated in Appendix \ref{app:ssh}.

\begin{figure}[b]
	\includegraphics[width=\linewidth]{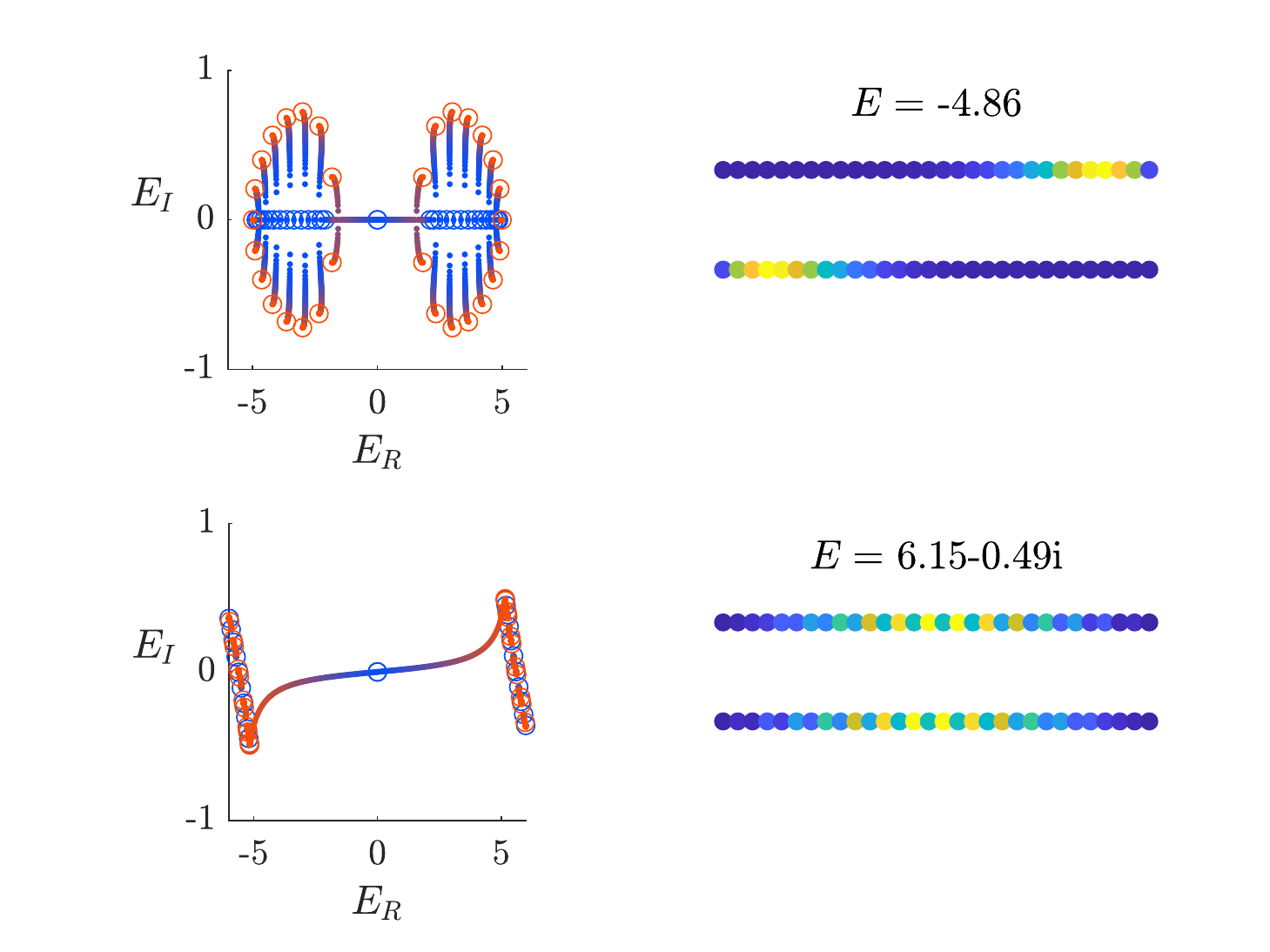}
	\caption{The left panel shows the eigenvalues of the SSH-chain with $N=30$ sites for $t_{l,1} = 1$, $t_{l,2} = 3$, $t_{r,1} = 2$ and $t_{r,2} = 4$ (upper left) and $t_{l,1} = i$, $t_{l,2} = 4$, $t_{r,1} = 0.5$ and $t_{r,2} = 8$ (lower left). Red circles correspond to periodic boundary conditions and blue circles to open boundary conditions. The dots show how the eigenvalues change when we change $\delta$ in steps of $0.01$ from $0$ to $1$. The right panel shows the right and left expectation values of the projection operator in a representative state for both sets of parameters in the upper right and lower right respectively.}
	\label{fig:ssh_figure}
\end{figure}

It has previously been observed \cite{YaWa2018,KuEdBuBe2018} that there is a pronounced skin effect for most parameter values, and that the eigenvalues are highly sensitive to changes in boundary conditions. This can be seen in the upper panel of Fig.~\ref{fig:ssh_figure}, where we to the left plotted the eigenvalues for a generic system for different values of $\delta$, and to the right an example of what the left and right eigenstates look like (for more details on this figure, see below).

In this section we study this sensitivity in more detail. Even though this system is not described by a Toeplitz matrix, and the method described in the beginning of this section at first glance thus seems to be unusable in this case, we notice that the square of the Hamiltonian is given by 
\begin{widetext}
\begin{equation}
	H_0^2 = \begin{pmatrix}
	t_{l,1}t_{r,1}+\delta^2t_{l,2}t_{r,2} & 0       & t_{l,1}t_{l,2}     & 0      & \dots   & 0        & t_{r,1}t_{r,2}\delta &0\\
	0           & t_{l,1}t_{r,1}+t_{l,2}t_{r,2}   & 0      & t_{l,1}t_{l,2}     &         &          &         & t_{r,1}t_{r,2}\delta\\
	t_{r,1}t_{r,2}          & 0       & \ddots & \ddots & \ddots  &          &         &0\\
	0           & t_{r,1}t_{r,2}      & \ddots & \ddots & \ddots  & \ddots   &         &\vdots\\
	\vdots      &         & \ddots & \ddots & \ddots  & \ddots   & t_{l,1}t_{l,2}      &0\\
	0           &         &        & \ddots & \ddots  & \ddots   & 0       & t_{l,1}t_{l,2}\\
	t_{l,1}t_{l,2}\delta     &         &        &        & t_{r,1}t_{r,2}      & 0        & t_{l,1}t_{r,1}+t_{l,2}t_{r,2}   & 0 \\
	0           & t_{l,1}t_{l,2}\delta & 0      & \dots  & 0       & t_{r,1}t_{r,2}       & 0       & t_{l,1}t_{r,1}+\delta^2 t_{l,2}t_{r,2}
	\end{pmatrix} \ .
\end{equation} 
\end{widetext}
This is almost a Toeplitz matrix with changes only in the parts of the matrix that would affect the boundary equations, so we can use the previously described method for $H^2_0$, and instead solve the problem $H^2_0\psi_{\alpha} = \lambda_{\alpha}\psi_{\alpha}$, with the appropriate boundary equations. The eigenvalues of $H^2_0$ will be the squares of the eigenvalues of $H_0$ (which come in $(\lambda,-\lambda)$ pairs when $N$ is even). Using the eigenvalues of the periodic SSH-chain together with the bulk equations of $H^2_0$, we get the following ansatz for the eigenvalues of $H^2_0$ and general $\delta$:
\begin{equation}
\label{eq:lambdasq}
	\lambda_{\alpha}^2 = t_{l,1}t_{r,1}+t_{l,2}t_{r,2}+t_{l,1}t_{l,2}e^{ i\alpha}+t_{r,1}t_{r,2}e^{- i\alpha}.
\end{equation}

The bulk equations now give us the following recurrence relation for the elements of $\psi_{\alpha}$:
\begin{align}
&t_{r,1}t_{r,2}\psi_{\alpha,n-2} + t_{l,1}t_{l,2}\psi_{\alpha,n+2} \nonumber\\
&-(t_{l,1}t_{l,2}e^{ i\alpha}+t_{r,1}t_{r,2}e^{- i\alpha})\psi_{\alpha,n} = 0 \ .
\end{align}
This is solved, for arbitrary integer $n$, by
\begin{align}
\psi_{\alpha,n} &= [c_1 + (-1)^n c_2] e^{ i\alpha n/2} \nonumber\\
&+ [c_3 +(-1)^n c_4] \Bigl( \frac{\sqrt{t_{r,1}}\sqrt{t_{r,2}}}{\sqrt{t_{l,1}}\sqrt{t_{l,2}}} \Bigr)^n e^{- i\alpha n/2},
\end{align}

where the constants $c_i$ are determined by the boundary equations
\begin{align}
(\delta^2-1) t_{l,2}t_{r,2}\psi_{\alpha,1} + t_{r,1}t_{r,2}(\delta \psi_{\alpha,N-1} - \psi_{\alpha,-1}) &= 0,\nonumber\\
t_{r,1}t_{r,2} (\delta\psi_{\alpha,N} - \psi_{\alpha,0})&= 0,\nonumber\\
t_{l,1}t_{l,2} (\delta \psi_{\alpha,1} - \psi_{\alpha,N+1}) &= 0,\nonumber\\
(\delta^2-1) t_{l,2}t_{r,2} \psi_{\alpha,N} + t_{l,1}t_{l,2}(\delta\psi_{\alpha,2} - \psi_{\alpha,N+2}) &=0 \ .
\end{align}
As for the Hatano-Nelson model, it turns out that it is convenient to apply a shift to $\alpha$
\begin{equation}
	\tilde{\alpha} = \alpha+ i\ln{\frac{\sqrt{t_{r,1}}\sqrt{t_{r,2}}}{\sqrt{t_{l,1}}\sqrt{t_{l,2}}}}.
\end{equation}

Then, the eigenvalues of $H$ are given by
\begin{align}
\label{eq:ssh_eigenvalues}
&\lambda_{\tilde{\alpha}} = \\
&\pm\sqrt{t_{l,1}t_{r,1}+t_{l,2}t_{r,2}+2\cos(\tilde{\alpha})\sqrt{t_{l,1}}\sqrt{t_{l,2}}\sqrt{t_{r,1}}\sqrt{t_{r,2}}} \nonumber,
\end{align}
where $\tilde{\alpha}$ follows from the determinant equation, which takes the form
\begin{align}
\label{eq:ssh_alphaeq}
&-\frac{\sin(\tilde{\alpha}(N/2+1))}{\sin(\tilde{\alpha})} +
(\delta^2 - 1)
\frac{
\sqrt{t_{l,2}}\sqrt{t_{r,2}}
}
{
\sqrt{t_{l,1}}\sqrt{t_{r,1}}
}
\frac{\sin(\tilde{\alpha}N/2)}{\sin(\tilde{\alpha})} \nonumber\\
&+ \delta^2 \frac{\sin(\tilde{\alpha}(N/2-1))}{\sin(\tilde{\alpha})} \nonumber\\
&+
\delta
\left[
\left(\frac{\sqrt{t_{r,1}}\sqrt{t_{r,2}}}{\sqrt{t_{l,1}}\sqrt{t_{l,2}}}\right)^{N/2}+
\left(\frac{\sqrt{t_{l,1}}\sqrt{t_{l,2}}}{\sqrt{t_{r,1}}\sqrt{t_{r,2}}}\right)^{N/2}
\right] = 0 \ .
\end{align}
We find that unless 
\begin{equation}
|t_{r,1}t_{r,2}| = |t_{l,1}t_{l,2}|,
\end{equation}
the spectrum has exponential sensitivity. In this special case, we have 
\begin{equation}
\frac{\sqrt{t_{l,1}}\sqrt{t_{l,2}}}{\sqrt{t_{r,1}}\sqrt{t_{r,2}}} =  e^{2\pi i\theta},
\end{equation}
for some $\theta\in[0,1]$.
As was the case for the HN-model, hermiticity is not required to get eigenvalues that are insensitive to the boundary conditions.
In Fig.~\ref{fig:ssh_figure}, we illustrate these results. In the upper left panel, we plot the eigenvalues of a system with $N=30$ sites, and hopping parameters
$t_{l,1} = 1, t_{r,1} = 2, t_{l,2} = 3, t_{r,2} = 4$, with $\delta$ varying from $\delta=0$ to $\delta = 1$.
We see that the system has a point gap, and that the spectrum is exponentially sensitive to the variation in $\delta$.
In the lower left panel of Fig.~\ref{fig:ssh_figure}, we plot the case $t_{l,1} = -i, t_{r,1} = 1/2, t_{l,2} = 4, t_{r,2} = 8$,
for which, on the contrary, the eigenvalues show a non-exponential sensitivity to the variation in $\delta$. In this case, the system has a line gap.

In the rightmost panel of Fig.~\ref{fig:ssh_figure}, we plot, for both sets of parameters, the left and right expectation values, $\langle \Pi_{\tilde\alpha,n} \rangle^{l,l} = | \psi^l_{\tilde\alpha,n}|^2$ and $\langle \Pi_{\tilde\alpha,n} \rangle^{r,r} = | \psi^r_{\tilde\alpha,n}|^2$, for a representative eigenstate,
showing a pronounced skin effect in the upper panel, while in the lower panel, the skin effect is absent.

As was the case for the HN-model, we observe a strong skin effect in the unbalanced case, which also exhibits
a point gap, and a strong $\delta$ dependence of the eigenvalues, in accordance with results in previous works.

For both sets of parameters in Fig.~\ref{fig:ssh_figure}, we observe a double zero-mode $E \approx 0$ for $\delta=0$,
which gradually merges with the rest of the spectrum upon varying $\delta$ from $\delta=0$ to $\delta=1$.
Even in the case $t_{l,1} = 1, t_{r,1} = 2, t_{l,2} = 3, t_{r,2} = 4$, where the rest of the spectrum is exponentially sensitive to varying $\delta$ away from zero, the energy of the $\delta = 0$ zero mode changes gradually with $\delta$. This corresponds to what was seen in e.g. \cite{BuBe2020}, where it was shown that the zero mode can be insensitive to boundary conditions even though the rest of the spectrum is not. In appendix \ref{app:ssh}, we obtain the condition for having a zero mode, which happens for
$\left| \frac{t_{l,2} t_{r,2}}{t_{l,1} t_{r,1}} \right| > 1$ in the large $N$ limit. This agrees with what was found in \cite{YaJiBe2022} and generalizes the condition found in e.g. \cite{KuEdBuBe2018}. We note that in the balancing condition, for which the spectrum is insensitive to changes in $\delta$, the left and right hopping parameters are `paired up', while in the condition for having a zero mode, the hopping parameters with index one and index two are `paired up'.

To make the system more useful for later, we add alternating on-site potentials according to
\begin{equation}
	H = H_0+V = \begin{pmatrix}
	v_1& t_{l,1} & & &&\delta t_{r,2}\\
	t_{r,1} &v_2 &t_{l,2}& &&\\
	& t_{r,2} &\ddots &t_{l,1}&&\\
	& & t_{r,1} & \ddots&\ddots&\\
	&&&\ddots&\ddots&t_{l,1}\\
	\delta t_{l,2}&&&&t_{r,1}&v_2
	\end{pmatrix} \ .
	\label{eq:ssh_hamiltonian_onsite}
\end{equation}
To find the eigenvalues, we note that 
\begin{equation}
H = H'+I\frac{v_1+v_2}{2},
\end{equation} 
where
\begin{equation}
H' = \begin{pmatrix}
v& t_{l,1} & & &&\delta t_{r,2}\\
t_{r,1} &-v &t_{l,2}& &&\\
& t_{r,2} &\ddots &t_{l,1}&&\\
& & t_{r,1} & \ddots&\ddots&\\
&&&\ddots&\ddots&t_{l,1}\\
\delta t_{l,2}&&&&t_{r,1}&-v
\end{pmatrix},
\end{equation}
and
\begin{equation}
v = \frac{v_1-v_2}{2}.
\end{equation}
Since
\begin{equation}
H'^2 = H_0^2+Iv^2,
\end{equation}
we can directly use the result from Eq.~\eqref{eq:lambdasq} and add $v^2$. The final result, then, is 
\begin{equation}
\label{eq:ssh_eigenvalues_potential}
\begin{split}
	&\lambda_{\tilde{\alpha}} = \frac{v_1+v_2}{2}\pm\\
	&\sqrt{v^2+t_{l,1}t_{r,1}+t_{l,2}t_{r,2}+2\cos(\tilde{\alpha})\sqrt{t_{l,1}}\sqrt{t_{l,2}}\sqrt{t_{r,1}}\sqrt{t_{r,2}}},
\end{split}
\end{equation}
where $\tilde\alpha$ is again given by Eq.~\eqref{eq:ssh_alphaeq}, and we see that alternating on-site potentials do not change the sensitivity of the eigenvalues to boundary conditions.

\subsection{Longer range hopping}

Both the Hatano-Nelson model and the SSH-chain are examples of models with nearest neighbor hopping. In this section we study some examples of models with longer range hopping and see if they can be balanced. In general, longer range hopping makes the recurrence relation in Eq.~\eqref{eq:recursive} more complicated, but when one restricts to the case of two parameters, the equations simplify.  

\subsubsection{Unidirectional hoppings}
First, we consider a system with nearest and next-nearest neighbour hopping to the left, but no hopping to the right. This is described by the matrix
\begin{equation}
H = \begin{pmatrix}
0          & t_l        & u_l      &                & \\
& \ddots   & \ddots & \ddots         & \\
&          & \ddots &  \ddots &u_l \\
\delta u_l   &          &        &\ddots &      t_l\\
\delta t_l   & \delta u_l &        &               & 0 
\end{pmatrix} \ .
\end{equation}
We note that for $\delta=0$, the matrix only has one eigenvector and all its eigenvalues are zero. 
For $\delta \neq 0$, we make the ansatz
\begin{equation}
\lambda_{\alpha} = t_le^{ i\alpha}+u_le^{2 i\alpha}. 
\end{equation}
The recurrence relation for the elements of $\psi_{\alpha}$ becomes
\begin{equation}\label{eq:recursive}
t_l\psi_{\alpha,n+1}+u_l\psi_{\alpha,n+2}-\lambda_{\alpha}\psi_{\alpha,n}=0,
\end{equation}
and is solved by
\begin{equation}
\psi_{\alpha,n} = \left(-\frac{t_l}{u_l}-e^{ i\alpha}\right)^n c_1+e^{i\alpha n}c_2.
\end{equation}
 The constants $c_1$ and $c_2$ are again found using the boundary equations
\begin{align}
\delta u \psi_{\alpha,1} - u \psi_{\alpha,N+1} &=0, \\
\delta t \psi_{\alpha,1} + \delta u \psi_{\alpha,2} - t \psi_{\alpha,N+1} - u \psi_{\alpha,N+2} &=0.
\end{align}
Dropping an unimportant overall factor, this gives us the following determinant equation for $\alpha$: 
\begin{equation}
\left(-\delta +e^{i \alpha N }\right)
 \left(\delta -\left[-\frac{t_l}{u_l} + e^{i \alpha}\right]^{N }\right) =0.
\end{equation}
One can use either factor to determine $\alpha$, both give rise to the same eigenvalues.
By using the first factor, one finds that the solutions for $\alpha$ only depend on $\delta$, not on
the other parameters of the model.
Parametrising $\delta = |\delta| e^{ i\phi}$, we obtain the following eigenvalues, with $j=0,1,\ldots,N-1$
\begin{equation}
\lambda_j = t_l |\delta|^{\frac{1}{N}} e^{ i\phi/N + 2 \pi  i j/N} +
u_l |\delta|^{\frac{2}{N}} e^{2i\phi/N + 4 \pi  i j/N}, 
\end{equation} 
which means that if $\delta$ is real and $\delta\in(0,1)$, the solutions will approach the solutions of the periodic case for large $N$. In the special case that $\delta=0$, all eigenvalues will be $0$ regardless of $N$.
This implies an extreme sensitivity to the boundary conditions for all values of $t_l$ and $u_l$, which makes sense, since there is no way to balance the hopping to the left with hopping in the opposite direction.
\begin{figure}
	\includegraphics[width=\linewidth]{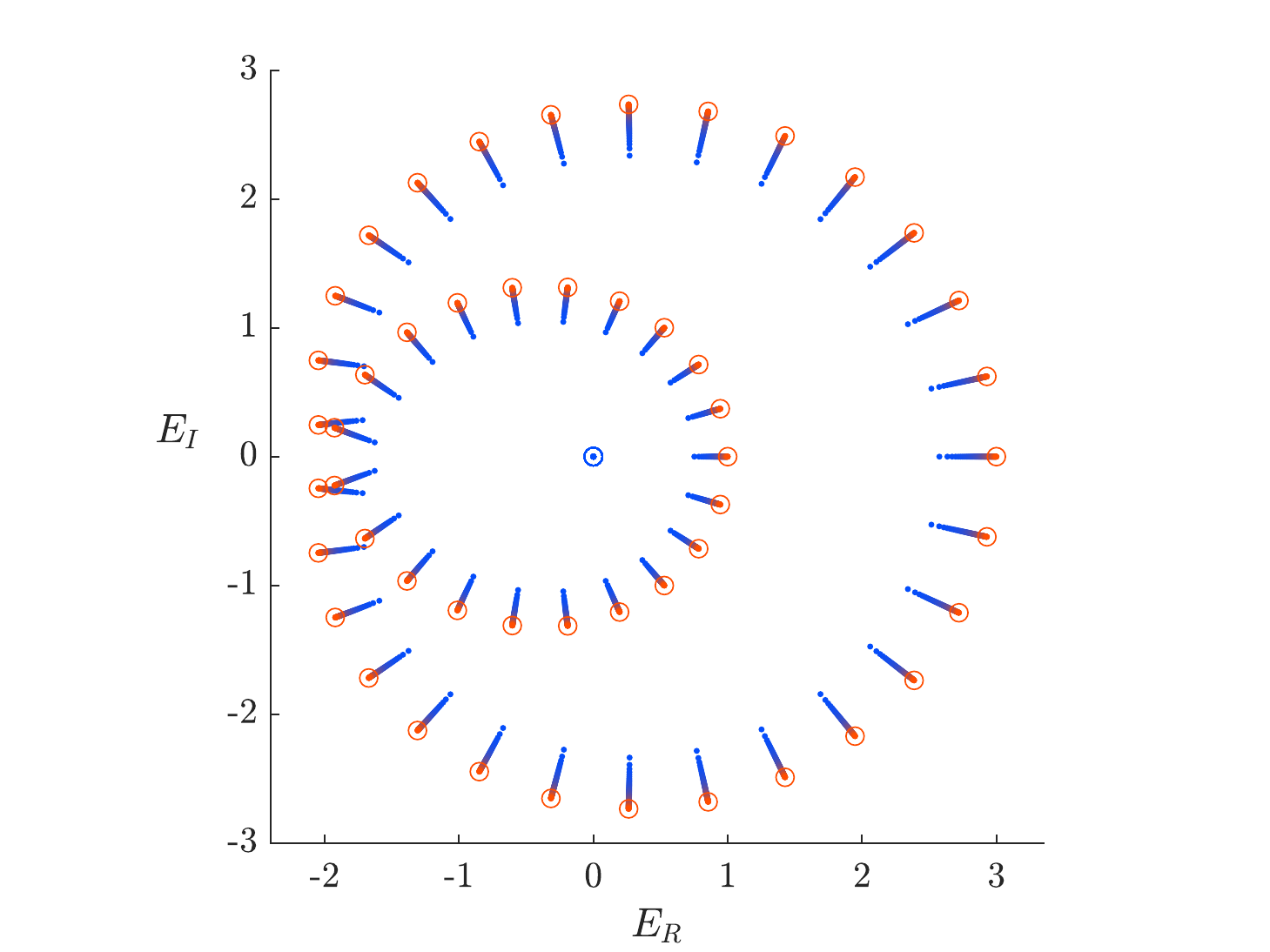}
	\caption{Eigenvalues for the unidirectional model plotted for $t_l = 1$, $u_l = 2$. Red circles correspond to periodic boundary conditions and blue circles to open boundary conditions. The dots show how the eigenvalues change when we change $\delta$ in steps of $0.01$ from $0$ to $1$. 
	}
	\label{fig:longrange_firstcase}
\end{figure}

In Fig.~\ref{fig:longrange_firstcase}, we plot the eigenvalues of a chain of length $N = 50$  with $t_l = 1$ and $u_l = 2$ for values of $\delta$ between $0$ and $1$, and see that even for very small values of $\delta$, the eigenvalues deviate a lot from $0$ and approach the eigenvalues of the periodic case.

\subsubsection{Hoppings in different directions}

Next, we study the Hamiltonian given by the matrix

\begin{figure*}
	\includegraphics[width=\linewidth]{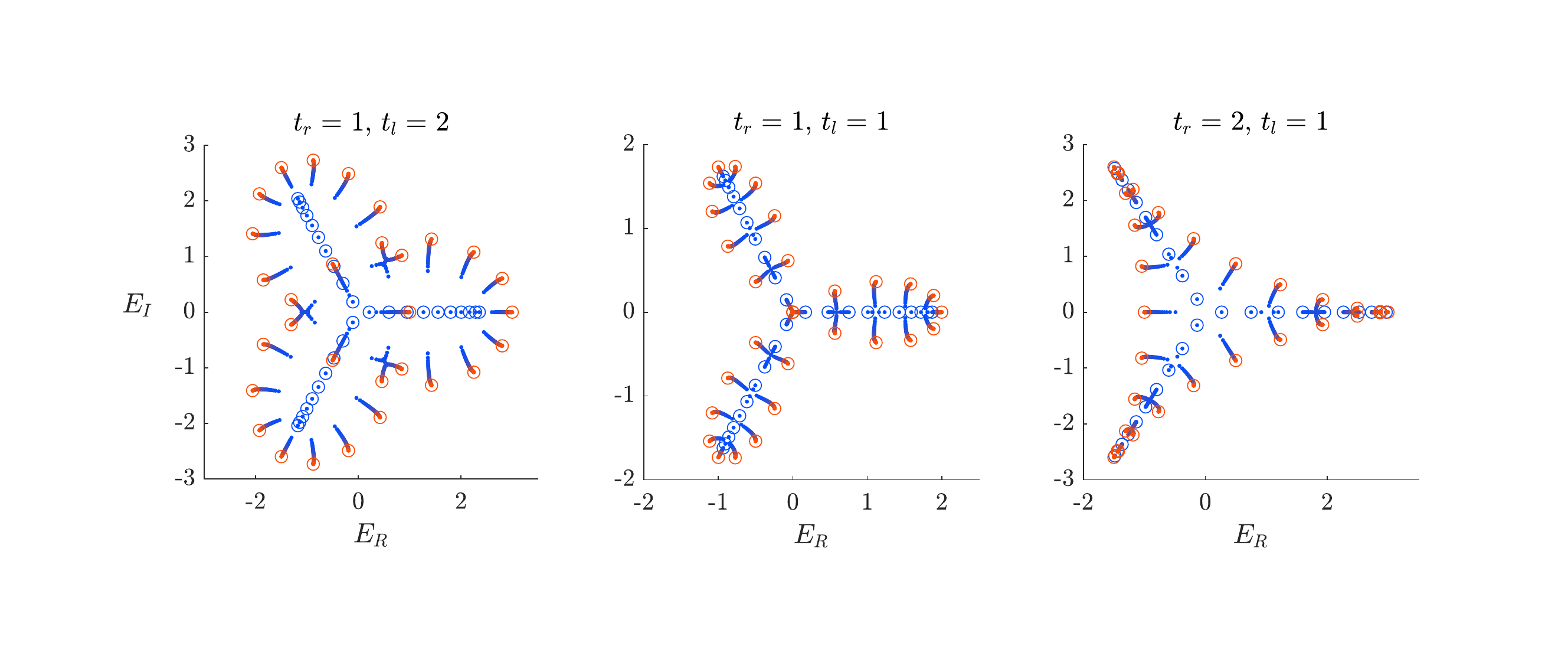}
	\caption{The eigenvalues of the chain with longer range hopping in different directions for different parameter values. Red circles correspond to periodic boundary conditions and blue circles to open boundary conditions. The dots show how the eigenvalues change when we change $\delta$ in steps of $0.01$ from $0$ to $1$. We note that the change in the eigenvalues when we go from $\delta= 0$ to $\delta = 0.01$ is much larger when $t_r\neq  u_l$. There is, however, still a significant change in the spectrum also for $t_r=u_l$. }
	\label{fig:longer_range_figure}
\end{figure*}

\begin{equation}
H = \begin{pmatrix}
0          & 0        &u_l       &         &\delta t_r \\
t_r          & \ddots   & \ddots & \ddots  & \\
           &  \ddots  & \ddots & \ddots  & u_l\\
\delta u_l   &    & \ddots & \ddots  & 0\\
 0          & \delta u_l &        & t_r       & 0 
\end{pmatrix},
\end{equation}
which describes a system with hoppings both to the right and to the left, but with different range.
We make the following ansatz for the eigenvalues:
\begin{equation}
	\lambda_{\alpha} = u_le^{2i\alpha}+t_re^{-i\alpha}.
\end{equation}
The recurrence relation for the elements of $\psi_{\alpha}$ becomes
\begin{equation}
	t_r\psi_{\alpha,n-1}+u_l\psi_{\alpha,n+2}-\lambda_{\alpha}\psi_{\alpha,n} = 0,
\end{equation}
which has the characteristic equation 
\begin{equation}
	x^3-e^{2i\alpha}\left(1+\frac{t_r}{u_l}e^{-3i\alpha}\right)x-\frac{t_r}{u_l} = 0.
\end{equation}
This equation has roots such that the elements of $\psi_{\alpha}$ are given by
\begin{equation}
\label{eq:longer-range2_eigenvectors}
	\psi_{\alpha,n} = c_1e^{ni\alpha}+c_2e^{-ni\alpha}x_+^n+c_3e^{-ni\alpha}x_-^n,
\end{equation}
where 
\begin{equation}
	x_{\pm} = \frac{1}{2}\pm\frac{1}{2}\sqrt{4r+1}\,\,\,\,\mathrm{with}\,\,\,\,r = \frac{t_r}{u_l}e^{-3i\alpha}.
\end{equation}
We note that 
\begin{equation}\label{eq:xpmrelations}
	x_+x_-  = -r\,\,\,\,\,\mathrm{and}\,\,\,\,\, x_{\pm}^2-x_{\pm}-r = 0.
\end{equation}
The constants $c_1,c_2,c_3$ are determined using the boundary equations, which in this case read
\begin{align}
& -t_r \psi_{\alpha,0} + \delta t_r \psi_{\alpha,N} = 0, \nonumber \\
& \delta u_l \psi_{\alpha,1} - u_l \psi_{\alpha,N+1} = 0, \nonumber \\
& \delta u_l \psi_{\alpha,2} - u_l \psi_{\alpha,N+2} = 0 \ .
\end{align}
The values of $\alpha$ for which this system of equations has solutions correspond to when the determinant of the system equals zero. The expression for the determinant is long, but can be simplified a lot using Eq.~\eqref{eq:xpmrelations}. Introducing the polynomial $p(n)$, recursively defined by 
\begin{equation}
	p(n) = -p(n-1)+rp(n-2),
\end{equation} 
with $p(0) = 0$ and $p(1) = -1$, the equation for $\alpha$ can be written in the following way:
\begin{multline}\label{eq:polynomialeq}
	-y^{3N+3}\left[(2+r)p(N)-2rp(N-1)+(-r)^{N+1}\right]\\
	\shoveleft{+\delta y^{2N+3}\left[2+2p(N)+2r(r-1)p(N-1)+ \right.}\\
	\shoveright{\left. +(2-r)y^{2N}(-r)^N\right]}\\
	\shoveleft{+\delta^2 y^{3N+3}\left[-y^{-2N}(2-r)+2rp(N)+ \right.}\\
	\shoveright{\left. +2r(1-r)p(N-1)-2(-r)^N\right]}\\
	-\delta^3 y^{2N+3}r\left[1+2p(N-1)+p(N)\right] = 0 \ ,
\end{multline}
where we introduced $y = e^{i\alpha}$ and we dropped a factor $(x_--x_+)$.
Though not manifestly so, the equation is proportional to $(r-2y^3)$ (we note that $r = t_r/(u_l y^3)$).
After dividing out by this factor, the remaining polynomial equation has degree $3N$ in terms of $y$, and
degree $N$ in terms of $t_r/u_l$. Thus, we find $3N$ solutions for $y$ (and hence $\alpha$), but they
come in triples which give rise to the same eigenvalue, so we obtain $N$ eigenvalues as wanted.

For $\delta= 0$, the equation reduces to
\begin{equation}
-y^{3N+3}\bigl((2+r)p(N)-2rp(N-1)+(-r)^{N+1}\bigr) = 0 
\end{equation}
and for $\delta= 1$, we have (up to unimportant factors)
\begin{equation}
\begin{split}
&y^{N+3} (y^N-1) \times\\
&\times\bigl( 1 - y^N p(N) + 2 r y^N p(N-1) + (-r)^N y^{2N} \bigr) = 0 \ .
\end{split}
\end{equation}

We note that Eq.~\eqref{eq:polynomialeq} contains terms with different powers of $r$. Since $r$ depends on $t_r/u_l$, one could expect the behavior of the eigenvalues to be insensitive to perturbations when $|u_l| = |t_r|$, but this is not the case for this model, in contrast to previous models. We can, however, see that the winding number of the Bloch Hamiltonian is non-zero when $|u_l|=|t_r|$, so the result is somewhat expected.
Technically, the reason for this behavior is that the eigenvectors Eq.~\eqref{eq:longer-range2_eigenvectors} generically have an exponential behavior, even when the parameters appear to be `balanced' as $|u_l|=|t_r|$.

We illustrate this in Fig.~\ref{fig:longer_range_figure}, where we plot the eigenvalues for a chain of length $N=30$, with parameters
$(u_l,t_r) = (2,1); (1,1); (1,2)$, and $\delta$ varying from $\delta = 0$ to $\delta = 1$.
We observe that in all three cases, the eigenvalues are sensitive to the boundary conditions, and there is a non-zero winding number for
$\delta = 1$, even in the case where the hopping to the left and right seem to be balanced.
Also the eigenstates have a pronounced skin effect (not shown).

\subsubsection{Example of a system not solvable by the method}\label{sec:unsolvable_system}

Since the described method to find eigenvalues relies on finding zeros of a polynomial equation of a degree proportional to the hopping range, it will typically fail for more complicated systems since we cannot solve such equations exactly. Nevertheless, we will briefly look at one such system, which will be of interest for later use.

\begin{figure}
	\includegraphics[scale=0.12]{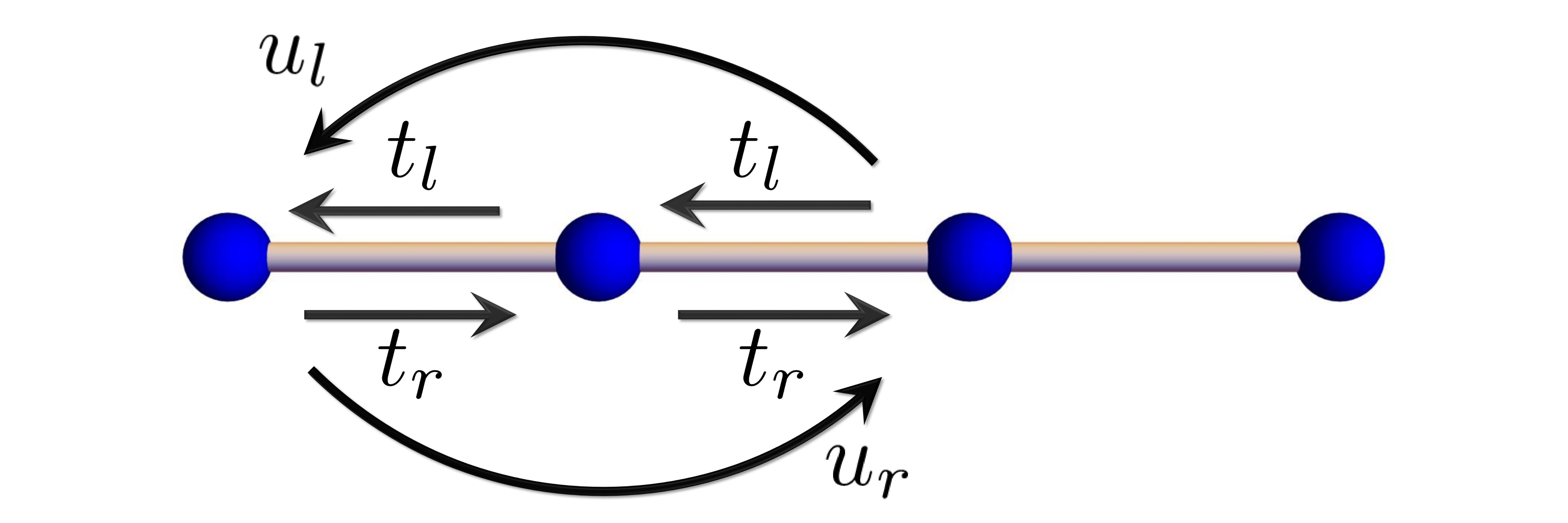}
	\caption{Model with longer range hopping that cannot be solved by the method. If $u_l = t_r$ and $u_r = t_l$, this is a chain of triangles.}
	\label{fig:longrange_unsolvable}
\end{figure}

We show this model in Fig.~\ref{fig:longrange_unsolvable}. The periodic version of this chain is described by the Bloch Hamiltonian
\begin{equation}
\label{eq:longer_range_bloch}
	H(k) = t_le^{ik}+t_re^{-ik}+u_le^{2ik}+u_re^{-2ik}.
\end{equation}
We first note that an interesting special case of the model described by Eq.~\eqref{eq:longer_range_bloch} is the case $u_l = t_r$ and $u_r = t_l$.
For these parameters, one ends up with a chain of triangles and a Bloch Hamiltonian given by
\begin{equation}\label{eq:small_triangle}
\begin{split}
	H(k) &= t_le^{ik}+t_re^{-ik}+t_re^{2ik}+t_le^{-2ik}\\
	&=\cos\left(\frac{3k}{2}\right)\left[t_re^{ik/2}+t_le^{-ik/2}\right].
\end{split}
\end{equation}
Unless $t_r/t_l$ is a phase, the eigenvalues in general form a loop in the complex plane, and the system will thus be exponentially sensitive to boundary conditions.

We note that there are several choices of the parameters, such that $H(k)$ in Eq.~\eqref{eq:longer_range_bloch} does not wind.

One way of achieving this is, is by setting
$t_l = t e^{i (\phi+\phi_1/2)}$, 
$t_r = t e^{i (\phi-\phi_1/2)}$, 
$u_l = u e^{i (\phi+\phi_2/2)}$, 
$u_r = u e^{i (\phi-\phi_2/2)}$.
For these parameters, the Hamiltonian becomes
\begin{equation}
H(k) = 2e^{i\phi}
\left[
t\cos(k+\phi_1/2)+u\cos(2k-\phi_2/2)
\right].
\end{equation}
Assuming that $t, u$ and the angles $\phi$ are real, this forms a straight line segment in the complex plane, and thus there is no winding of the spectrum in this case, which means that for these parameter values the spectrum has non-exponential sensitivity. We note that the condition we get on the hopping parameters for this to be the case, is slightly different from previous conditions in systems with nearest neighbor hopping. Namely, we get a conditions also on the relative phases of the parameters, and not only on the absolute values.

A second way of achieving this, is the case
$t_l = t e^{i \phi_1}$
$u_l = t e^{i \phi_1 + i \phi}$
$t_r = t e^{i \phi_2}$
$u_R = t e^{i \phi_2 - i \phi}$.
For these parameters, the Hamiltonian becomes
\begin{equation}
H(k) =
4 t e^{i(\phi_1+\phi_2)/2} \cos((k+\phi)/2)
\cos((3k + \phi + \phi_1 - \phi_2)/2) \ .
\end{equation}
Again, for $t$ and the phases $\phi$ real, this forms a straight line segment in the complex plane, implying that the spectrum has non-exponential sensitivity.

Finally, there is a third case,
$t_l = t e^{i \phi_1}$
$t_r = t e^{i \phi_1 + \phi}$
$u_l = t e^{i \phi_2}$
$u_R = t e^{i \phi_2 +2 i \phi}$.
In this case, the Hamiltonian becomes
\begin{equation}
H(k) =
2 (
t e^{i \phi_1 + \phi/2} \cos(k-\phi/2)
+
u e^{i \phi_2 + \phi} \cos(2k - \phi)
) \ .
\end{equation}
In this case, the values of $H(k)$ (with $t, u$ and the phases $\phi$ real) do not generically form a straight line in the complex plane.
However, because the argument of the second cosine is twice the argument of the first cosine, one obtains a curve where the part with $\pi \leq k \leq 2\pi$ traces back the part of the curve with $0 \leq k \leq \pi$.
Again, we conclude that there is no winding for these parameters, and hence the spectrum has non-exponential sensitivity.

We believe that the three cases above exhaust the ways in which one can obtain an $H(k)$ that does not wind.

\section{Towards two-dimensional systems}
\label{sec:two-dimensional-systems}

In this section, we consider two-dimensional systems that we construct by stacking one-dimensional chains studied in Sec.~\ref{sec:one-dimensioal-systems}. Contrary to the one-dimensional case, we will here allow for more complicated boundary conditions, so in the direction of stacking, we will have two parameters, $\delta_2$ and $\delta_2'$ determining the boundary conditions.
The one-dimensional chains, however, have the regular coupling by a single parameter $\delta_1$, which takes values in the interval $[0,1]$. Suppose we stack $N_2$ chains, each containing $N_1$ sites. The Hamiltonian for such a system can be described by the $N_1N_2\times N_1N_2$-matrix

\begin{equation}
H = \begin{pmatrix}
A(\delta_1) & B(\delta_1) & & \delta_2 C(\delta_1)\\
C(\delta_1) & A(\delta_1) & \ddots &\\
& \ddots & \ddots & B(\delta_1)\\
\delta_2' B(\delta_1)  & & C(\delta_1)& A(\delta_1)
\end{pmatrix},
\end{equation}
where $A(\delta_1)$, $B(\delta_1)$ and $C(\delta_1)$ are $N_1\times N_1$-matrices describing the one-dimensional chains and the coupling between them respectively (we drop the dependence of $A$, $B$, and $C$ on the other parameters of the model)
\footnote{%
In principle, it is typically not too complicated to consider two in-equivalent boundary conditions $\delta_1$ and $\delta_1'$ in the chains that are stacked. For instance, we do this for the HN-model in App.~\ref{app:hn-more-general}. Because this leads to more cumbersome notation $A(\delta_1,\delta_1')$, etc., we do not consider this possibility explicitly here.
}.

To find analytic expressions for this kind of matrix for arbitrary $\delta_1$ and $\delta_2,\delta_2'$ is in general a very hard problem. In particular it is difficult to find the eigenvalues for open boundary conditions in both directions, i.e. for all $\delta$'s being zero. If one were to try to implement the method used for one-dimensional systems, by considering a unit cell of size $N_2$, one would end up with several bulk equations that depend on the size of the system, and this would not be solvable.
Alternatively, one can simplify the bulk equations, by making use of the two-dimensional nature of the problem, and write the elements of the eigenvectors as $\psi_{i,j}$. This indeed leads to a simple bulk equation, but also results in the complication that the boundary equations in one direction should be satisfied for all possible locations in the other direction. It turns out that this approach does not work either.

Below, we consider two special types of boundary conditions, for which we can obtain an analytic solution, but we start even simpler, by briefly noting that one can construct solvable models on the square lattice (for details, see App.~\ref{app:trival-square-lattice}).
To do this, we simply take a one-dimensional model that we solved, i.e., for which we know the functional form of the eigenvalues $\lambda_\alpha$, as well as the equation for $\alpha$, denoted by ${\rm eq}_\alpha$.
The square lattice model now has the hopping parameters of this model in the horizontal direction, and the hopping parameters of a second solved model in the vertical direction, with eigenvalues $\lambda_\beta$, where $\beta$ satisfies ${\rm eq}_\beta$. The eigenvalues of the square lattice model are then simply given by $\lambda_{\alpha,\beta} = \lambda_{\alpha} + \lambda_{\beta}$, since the two directions are completely independent of each other.

We will, however, be interested in studying more complicated models that cannot be solved exactly if we have open boundary conditions. Thus, in this section, we will study two special cases of boundary conditions that \textit{can} be solved analytically and then use them to try and make predictions of the system with open boundary conditions.

\textit{BC 1.} We will begin by assuming that $\delta_2 = \delta_2'= 1$, making the system periodic in one direction, while we can interpolate between open and periodic boundary conditions in the other direction. Taking the Fourier transform of $H$ in the periodic direction, we get a Bloch Hamiltonian of the form

\begin{equation}
	\tilde{H}_j= A(\delta_1) + \omega_j B(\delta_1) + \omega_j^{-1}C(\delta_1),
	\label{eq:smallmatrix}
\end{equation} 
where now $\omega_j = \exp(2\pi i j/N_2)$, which has the property that if $\lambda_{jk}$ is an eigenvalue of $\tilde{H}_j$ with corresponding eigenvector $v_{jk}$,
then $\lambda_{jk}$ is an eigenvalue of $H$ with corresponding eigenvector $\begin{pmatrix}
v_{jk}&v_{jk}\omega_j&v_{jk}\omega_j^2&\dots&v_{jk}\omega_j^{N_2-1}
\end{pmatrix}^T$.

\textit{BC 2.} Next, we consider the case $\delta_2' = \delta_2^{-1}$. Define the matrix
\begin{equation}
    T = \begin{pmatrix}
    \delta_2^{1/N_2}& & &\\
    & \delta_2^{2/N_2} &&\\
    && \ddots&\\
    &&&\delta_2
    \end{pmatrix}\otimes I_{N_1}.
\end{equation}
Then
\begin{equation}
    T^{-1}HT = \begin{pmatrix}
A & B\delta_2^{1/N_2} & & C\delta_2^{-1/N_2} \\
C\delta_2^{-1/N_2} & A & \ddots &\\
& \ddots & \ddots & B\delta_2^{1/N_2} \\
 B\delta_2^{1/N_2}  & & C\delta_2^{-1/N_2} & A
\end{pmatrix}.
\end{equation}

This describes a system with periodic boundary conditions in one direction, and its eigenvalues are given by the eigenvalues of the matrices 
\begin{equation}
	\tilde{H}_j= A(\delta_1) + \omega_j\delta_2^{1/N_2} B(\delta_1) + \omega_j^{-1}\delta_2^{-1/N_2}C(\delta_1) \ ,
\end{equation} 
where $\omega_j = \exp(2\pi ij /N_2)$. For $\delta_2 = 1$, this case reduces to \textit{BC 1} as it should.

\subsection{Stacking Hatano-Nelson chains}

\begin{figure}
	\includegraphics[scale=0.12]{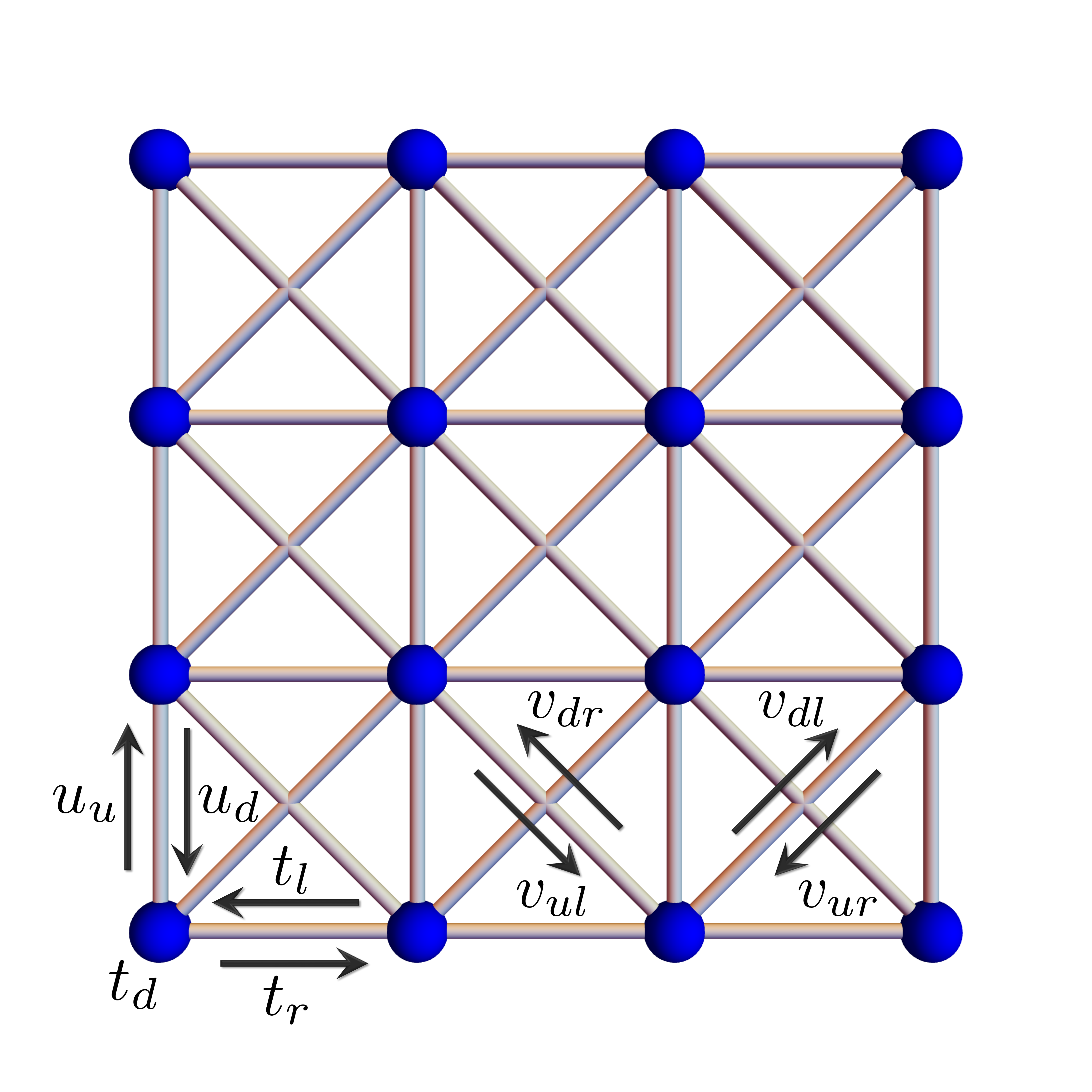}
	\caption{The lattice consisting of stacked Hatano-Nelson chains.}
	\label{fig:lattice_1}
\end{figure}

Now we study an explicit model, which is insensitive to boundary conditions, for correctly chosen parameters. 
Consider the lattice shown in Fig.~\ref{fig:lattice_1} with boundary conditions according to \textit{BC 1}. 
In this system, we have 
\begin{equation}
A(\delta_1) = \begin{pmatrix}
t_d  & t_l &        &        & \delta_1t_r\\
t_r & t_d  & t_l     &        &   \\
& t_r & \ddots & \ddots &   \\
&    & \ddots & \ddots & t_l\\
\delta_1t_l &    &        &  t_r    & t_d
\end{pmatrix},
\end{equation}

\begin{equation}
B(\delta_1) = \begin{pmatrix}
u_d  & v_{dl} &        &        & \delta_1v_{dr}\\
v_{dr} & u_d  & v_{dl}     &        &   \\
& v_{dr} & \ddots & \ddots &   \\
&    & \ddots & \ddots & v_{dl}\\
\delta_1v_{dl} &    &        &  v_{dr}    & u_d
\end{pmatrix},
\end{equation}
and
\begin{equation}
C(\delta_1) = \begin{pmatrix}
u_u  & v_{ul} &        &        & \delta_1v_{ur}\\
v_{ur} & u_u  & v_{ul}     &        &   \\
& v_{ur} & \ddots & \ddots &   \\
&    & \ddots & \ddots & v_{ul}\\
\delta_1v_{ul} &    &        &  v_{ur}    & u_u
\end{pmatrix}.
\end{equation}

The eigenvalues of the matrix $H$ can be found by diagonalizing the matrix $\tilde{H}_j$ defined in Eq.~\eqref{eq:smallmatrix}. In this case, we have
\begin{equation}
	\tilde{H}_j = \begin{pmatrix}
	h_d  & h_l &        &        & \delta_1h_r\\
	h_r & h_d   & h_l     &        &   \\
	& h_r & \ddots & \ddots &   \\
	&    & \ddots & \ddots & h_l\\
	\delta_1h_l &    &        &  h_r     & h_d 
	\end{pmatrix},
\end{equation}
where
\begin{equation}
	\begin{split}
	h_d &= t_d+\omega_ju_d+\omega_j^{-1}u_u,\\
	h_r &= t_r+\omega_jv_{dr}+\omega_j^{-1}v_{ur},\\
	h_l &= t_l+\omega_jv_{dl}+\omega_j^{-1}v_{ul}.
	\end{split}
\end{equation}
	\begin{figure*}[t]
		\includegraphics[width=\linewidth]{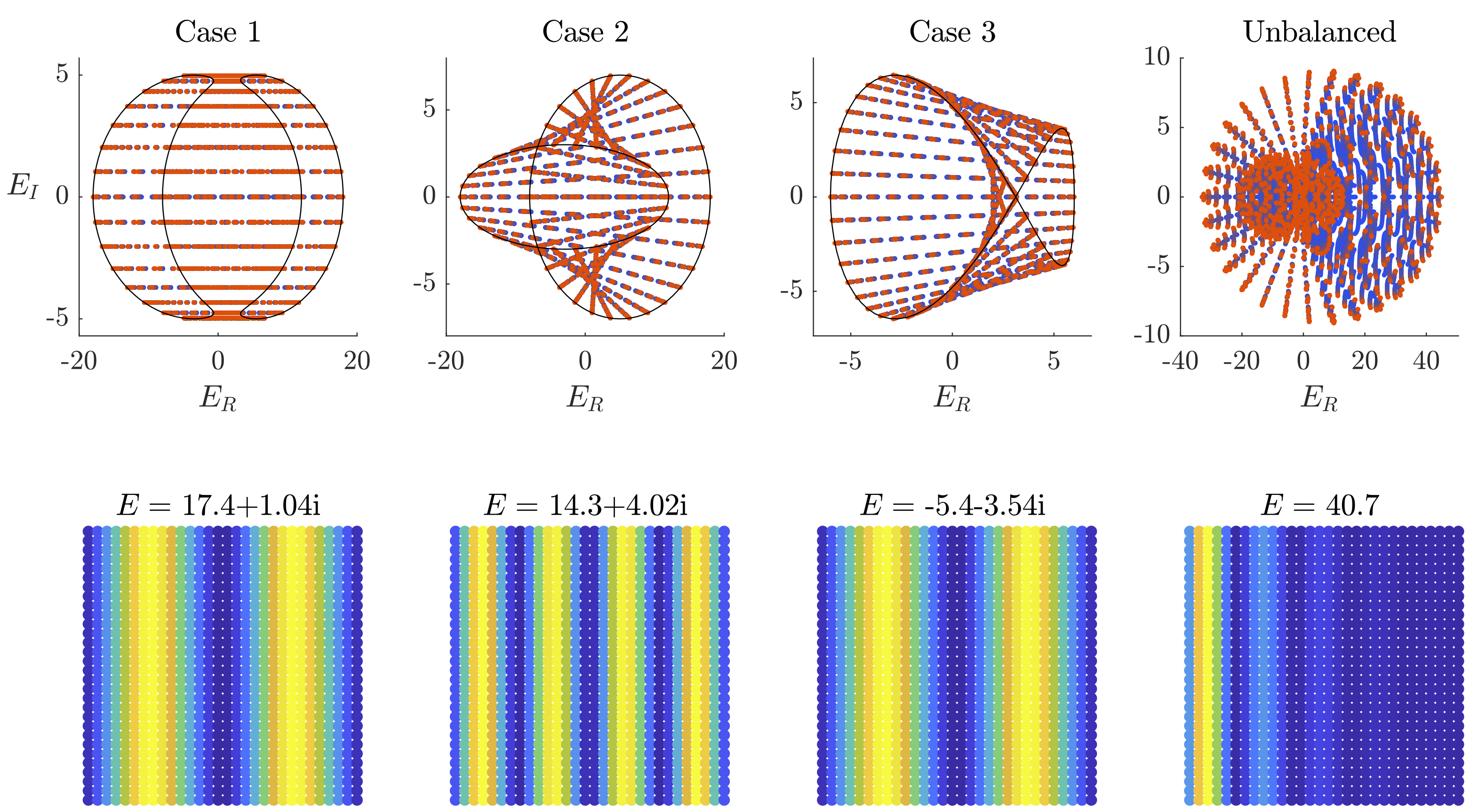}
		\caption{
		Plots for the stacked HN-model, with system sizes $N_1 = N_2 = 30$. 
		Upper panel: eigenvalues of systems representative of the different cases, including one unbalanced case. The dots show how the eigenvalues change when we change $\delta$ in steps of $0.01$ from $0$ (blue dots) to $1$ (red dots).  We also plot the curves $z_{\pm}(t)$ and see that the eigenvalues lie on straight lines connecting them. Lower panel: right expectation values of a representative state for each of the respective cases. For the parameter values in the four cases, we refer to the main text.
		}
		\label{fig:stacked_hn_cases}
	\end{figure*}
The matrix $\tilde{H}_j$ is precisely of the form discussed in Eq.~\eqref{eq:very_simple_hamiltonian} and its eigenvalues are thus given by
\begin{equation}\label{eq:squareevals}
	\lambda_{j\alpha} = h_d+2\sqrt{h_l}\sqrt{h_r}\cos(\tilde{\alpha}),
\end{equation}
where $\tilde{\alpha}$ is determined by the equation
\begin{equation}
	\begin{split}
	&\delta_1\left(\frac{h_l^{N_1/2}}{h_r^{N_1/2}}+\frac{h_r^{N_1/2}}{h_l^{N_1/2}}\right)\\&-\frac{\sin[(N_1+1)\tilde{\alpha}]}{\sin(\tilde{\alpha})}+\delta_1^2\frac{\sin[(N_1-1)\tilde{\alpha}]}{\sin(\tilde{\alpha})} = 0.
	\end{split}
\end{equation}
We note that for
\begin{equation}\label{eq:square_condition}
	 \left|\frac{h_l}{h_r}\right| = \left|\frac{t_l+\omega_jv_{dl}+\omega_j^{-1}v_{ul}}{t_r+\omega_jv_{dr}+\omega_j^{-1}v_{ur}}\right|= 1, 
\end{equation}
the spectrum has non-exponential sensitivity. This condition is independent of $t_d, u_d$ and $u_u$, which is reasonable since these parameters are either on-site potentials or hoppings in the direction of periodic boundary conditions. In case Eq.~\eqref{eq:square_condition} is fulfilled, as was previously shown, $\tilde{\alpha}$ is real and thus $\cos(\tilde{\alpha})\in[-1,1]$. This means that the eigenvalues given in Eq.~\eqref{eq:squareevals} are contained in the set given by the values of the function
\begin{equation}
f(t,c) = z_1(t)+cz_2(t),
\end{equation} 
where
\begin{equation}
z_1(t) = t_d+e^{it}u_d+e^{-it}u_u,
\end{equation}
and
\begin{equation}
z_2(t) = 2\sqrt{t_r+e^{it}v_{dr}+e^{-it}v_{ur}}\sqrt{t_l+e^{it}v_{dl}+e^{-it}v_{ul}},
\end{equation}
with $c\in[-1,1]$ and $t \in [0,2\pi]$.
We obtained $z_1(t)$ and $z_2(t)$ from $h_d$ and $2 \sqrt{h_l}\sqrt{h_r}$ respectively by replacing $\omega_j$ with $e^{it}$.
We note that for each $t$, the values of $f(t,c)$ form a line segment in the complex plane between the points $z_{\pm}(t) = z_1(t)\pm z_{2}(t)$. That is,
\begin{equation}
\begin{split}
z_{\pm}(t) &= t_d+e^{it}u_u+e^{-it}u_d \\&\pm  2\sqrt{t_r+e^{it}v_{ur}+e^{-it}v_{dr}}\sqrt{t_l+e^{it}v_{ul}+e^{-it}v_{dl}}
\end{split}
\end{equation}
where $t\in[0,2\pi]$. This means that irrespective of the value of $\delta_1$, the eigenvalues of $H$ will be contained in the area formed by connecting the points $z_{\pm} (t)$ on the two curves by straight lines. We also note that as we increase $N_1$, the number of different solutions to the equation for $\tilde{\alpha}$ will increase, and as we increase $N_2$ the number of $\omega_j$ increases, and thus, as we make the system larger, more and more of the area will be covered by eigenvalues. Therefore, as we will see is useful later, we argue that in the thermodynamic limit, the spectrum in this area will be dense.

We are interested in finding out when the parameter values fulfill Eq.~\eqref{eq:square_condition}. This happens when $h_l = e^{2\pi ir/N_2}h_r$, which means that
\begin{equation}
	t_l+\omega_jv_{dl}+\omega_j^{-1}v_{ul} = e^{2\pi ir/N_2}(t_r+\omega_jv_{dr}+\omega_j^{-1}v_{ur}),
\end{equation}
where $r$ is a real number. Here we will study some particularly interesting cases. From now on we will assume that the parameters of $H$ are real. 

\vspace{0.5cm}
 \textit{Case 1.} For $h_r = h_l^*$, Eq.~\eqref{eq:square_condition} is fulfilled and we have $r = -\arg(h_r)N_2/\pi$. Furthermore, we get  $t_r = t_l$, $v_{ur} = v_{dl}$ and $v_{dr} = v_{ul}$. We note that the Hermitian case is a special case of this. In the upper panel of Fig.~\ref{fig:stacked_hn_cases}, we plot the curves $z_{\pm}(t)$ together with the eigenvalues of a system with $t_r = t_l = 2$, $v_{ur} = v_{dl} = 4$, $v_{dr} = v_{ul} = 3$, $t_d = 1$, $u_u = -3$ and $u_d = 2$ for $N_1 = N_2 = 30$ when we change $\delta_1$ in steps of $0.01$ from $\delta_1 = 0$ (blue dots) to $\delta_1 = 1$ (red dots). We see that the eigenvalues lie along straight line segments connecting the two curves and that the spectrum has non-exponential sensitivity. In the lower panel, we plot the right expectation value $\langle \Pi_{\tilde\alpha,n} \rangle^{r,r} = | \psi^r_{\tilde\alpha,n}|^2$ for a representative eigenvector, which shows that we have no skin effect at these parameter values.

\vspace{0.5cm}
\textit{Case 2.} The case $r = 0$ implies $h_r = h_l$, i.e. $t_l = t_r, v_{ul} = v_{ur}, v_{dl} = v_{dr}$, which means that the three matrices $A$, $B$ and $C$ are Hermitian, while the full Hamiltonian $H$ might not be. The eigenvalues are in this case given by
	\begin{equation}
		\lambda_{j\tilde{\alpha}} = h_d+2h_r\cos(\tilde{\alpha}),
	\end{equation}
	where $\tilde{\alpha}$ is given by the equation
	\begin{equation}	
		2\delta_1-\frac{\sin[(N_1+1)\tilde{\alpha}]}{\sin(\tilde{\alpha})}+\delta_1^2\frac{\sin[(N_1-1)\tilde{\alpha}]}{\sin(\tilde{\alpha})} = 0.
	\end{equation}
	In this case, $z_{\pm}(t)$ reduces to
	\begin{equation}
		z_{\pm}(t) = t_d \pm 2t_r +  e^{it}(u_d\pm 2v_{dr}) + e^{-it}(u_u\pm v_{ur}),
	\end{equation}
	where $t\in[0,2\pi]$. That is, two ellipses centered at $t_d\pm 2t_r$. In the upper panel of Fig.~\ref{fig:stacked_hn_cases}, we plot the curves $z_{\pm}(t)$ together with the eigenvalues of a system with $t_r = t_l = 2$, $v_{ur} = v_{ul} = 3$, $v_{dr} = v_{dl} = 4$, $t_d = 1$, $u_u = -3$ and $u_d = 2$ for $N_1 = N_2 = 30$ when we change $\delta_1$ in steps of $0.01$ from $\delta_1 = 0$ (blue dots) to $\delta_1 = 1$ (red dots). In the lower panel, we plot the right expectation value $\langle \Pi_{\tilde\alpha,n} \rangle^{r,r} = | \psi^r_{\tilde\alpha,n}|^2$ for a representative eigenvector, which shows that we have no skin effect at these parameter values.

\vspace{0.5cm}	
\textit{Case 3.} The case $r = j$ (i.e., $e^{2 \pi i r/N_2} = \omega_j$) implies that $t_l = v_{ur}, v_{dl} = t_r, v_{ul} = v_{dr} = 0$. This is a non-Hermitian matrix with eigenvalues given by
	\begin{equation}\label{eq:square_evals}
	\begin{split}
		\lambda_{j\tilde{\alpha}} &= t_d+\omega_ju_d+\omega_j^{-1}u_u\\&+2\cos(\tilde{\alpha})\sqrt{t_l+\omega_jt_r}\sqrt{\omega_j^{-1}(t_l+\omega_jt_r)},
	\end{split}
	\end{equation}
	where $\tilde{\alpha}$ is given by the equation
	\begin{equation}
	\begin{split}
		\delta_1^2\sin[\tilde{\alpha}(N_1-1)]&+2\delta_1\cos\left(\pi j\frac{N_1}{N_2}\right)\sin(\tilde{\alpha})\\&-\sin[\tilde{\alpha}(N_1+1)] =0.
	\end{split}
	\end{equation}

	Rewriting the eigenvalues as
	\begin{equation}
	\lambda_{j\tilde{\alpha}} = t_d+\omega_ju_d+\omega_j^{-1}u_u+2\cos(\tilde{\alpha})(t_le^{-\pi ij/N_1}+t_re^{\pi ij/N_1}),
	\end{equation}
	shows that $z_{\pm}(t)$ in this case reduce to
	\begin{equation}
		z_{\pm}(t) = t_d+u_de^{it}+u_ue^{-it}\pm  2(t_le^{-it/2}+t_re^{it/2}),
	\end{equation}
	where $t\in[0,2\pi]$. These two curves are actually two segments that together form a closed loop $s(t')$ in the complex plane, described by
	\begin{equation}\label{eq:parametrized_curve}
		\begin{cases}
		\mathrm{Re}[s(t')] = t_d+(u_u+u_d)\cos(2t')+2(t_l+t_r)\cos(t'),\\
		\mathrm{Im}[s(t')] = (u_d-u_u)\sin(2t')+2(t_r-t_l)\sin(t'),
		\end{cases}
	\end{equation}
	where $t'\in[0,2\pi]$. The eigenvalues of $H$ will thus lie on line segments joining two points on this curve. 
	
	In the upper panel of Fig.~\ref{fig:stacked_hn_cases}, we plot the curves $z_{\pm}(t)$ together with the eigenvalues of a system with $t_r = v_{dl} = 1$, $t_l =  v_{ur} = 2$, $v_{dr} = v_{ul} = 0$, $t_d = 1$, $u_u = -3$ and $u_d = 2$ for $N_1 = N_2 = 30$ when we change $\delta_1$ in steps of $0.01$ from $\delta_1 = 0$ (blue dots) to $\delta_1 = 1$ (red dots). In the lower panel, we plot the right expectation value $\langle \Pi_{\tilde\alpha,n} \rangle^{r,r} = | \psi^r_{\tilde\alpha,n}|^2$ for a representative eigenvector, which shows that we have no skin effect at these parameter values.

\vspace{0.5cm}	
\textit{Case 4.} The case $r = -j$ (i.e., $e^{2\pi i r/N_2} = \omega_j^{-1}$) implies that $t_l = v_{dr},v_{dl} = v_{ur} = 0,v_{ul} = t_r$. This case is similar to the previous one, and therefore not shown in Fig.~\ref{fig:stacked_hn_cases}.

\vspace{0.5cm}	
\textit{Unbalanced.}
Finally, we consider an unbalanced case, with parameters $t_d = 1$, $t_r = 2$, $t_l = 3$, $u_u = 4$, $v_{ur} = 5$, $v_{ul} = 6$, $u_d = 7$, $v_{dr} = 8$, $v_{dl} = 9$. From the plot in the upper panel of Fig.~\ref{fig:stacked_hn_cases}, where we have plotted the eigenvalues of this system for $N_1 = N_2 = 30$ when changing $\delta_1$ in steps of $0.01$ from $\delta_1 = 0$ (blue dots) to $\delta_1 = 1$ (red dots), it is clear that the eigenvalues do not lie on straight lines between two curves.
In the lower panel, we plot the right expectation value $\langle \Pi_{\tilde\alpha,n} \rangle^{r,r} = | \psi^r_{\tilde\alpha,n}|^2$ for a representative eigenvector, which shows that in this case, we do have a skin effect.

\vspace{0.5cm}	

Comparing the plots in Fig.~\ref{fig:stacked_hn_cases}, we see a clear difference in behavior of the eigenvalues between the balanced and unbalanced cases. Similar to what we saw for the Hatano-Nelson model and the SSH-chain, we see that the eigenvalues of the unbalanced system change significantly even for a very small change in $\delta_1$, while there is a much less drastic change for the balanced system.

\subsubsection{Special case: A triangular lattice}

\begin{figure}[t]
	\includegraphics[scale=0.125]{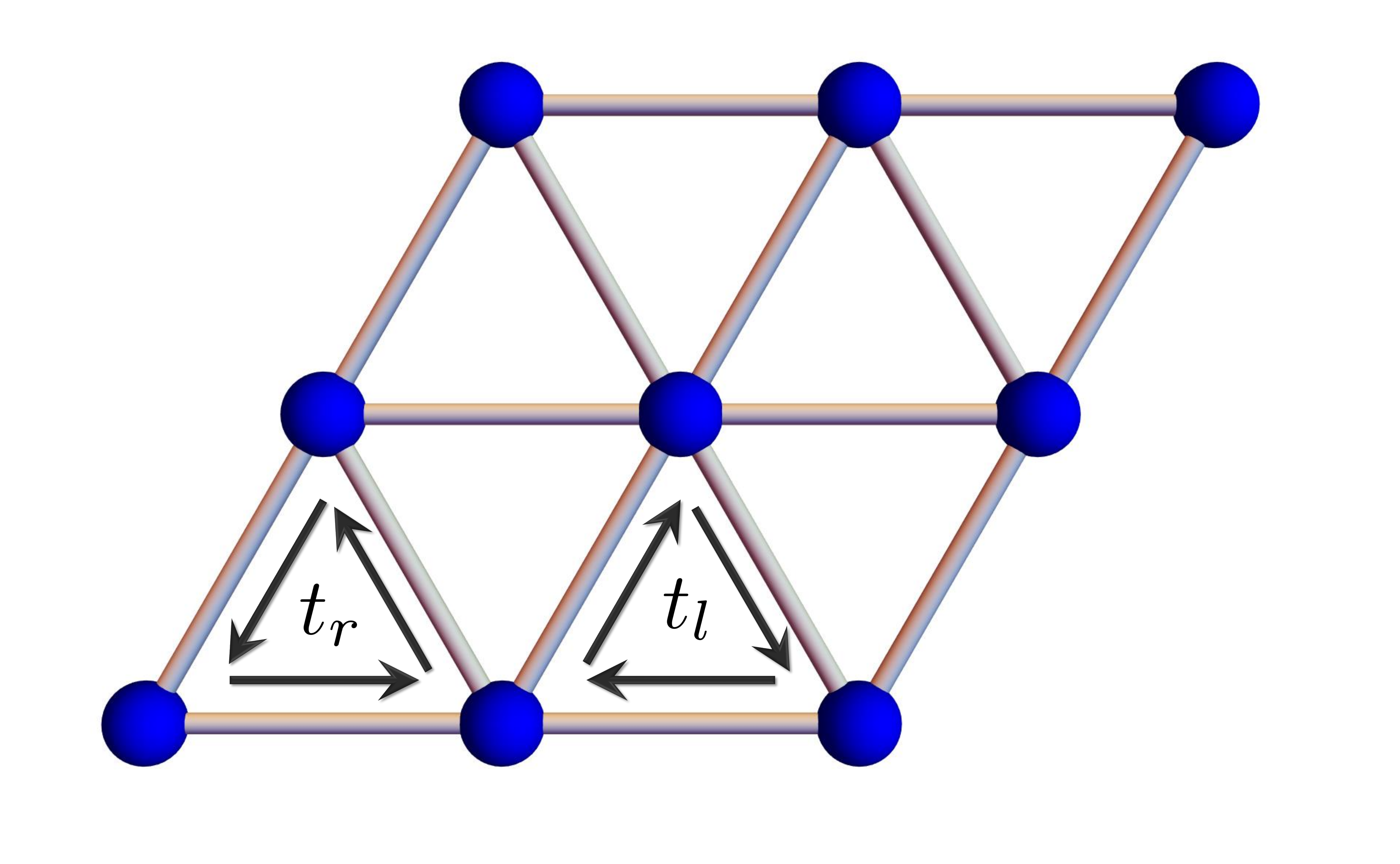}
	\caption{The triangular lattice resulting from implementing the balancing conditions from \emph{Case 3} in both directions.}
	\label{fig:triangular_lattice}
\end{figure}

\begin{figure}[t]
	\includegraphics[width=\linewidth]{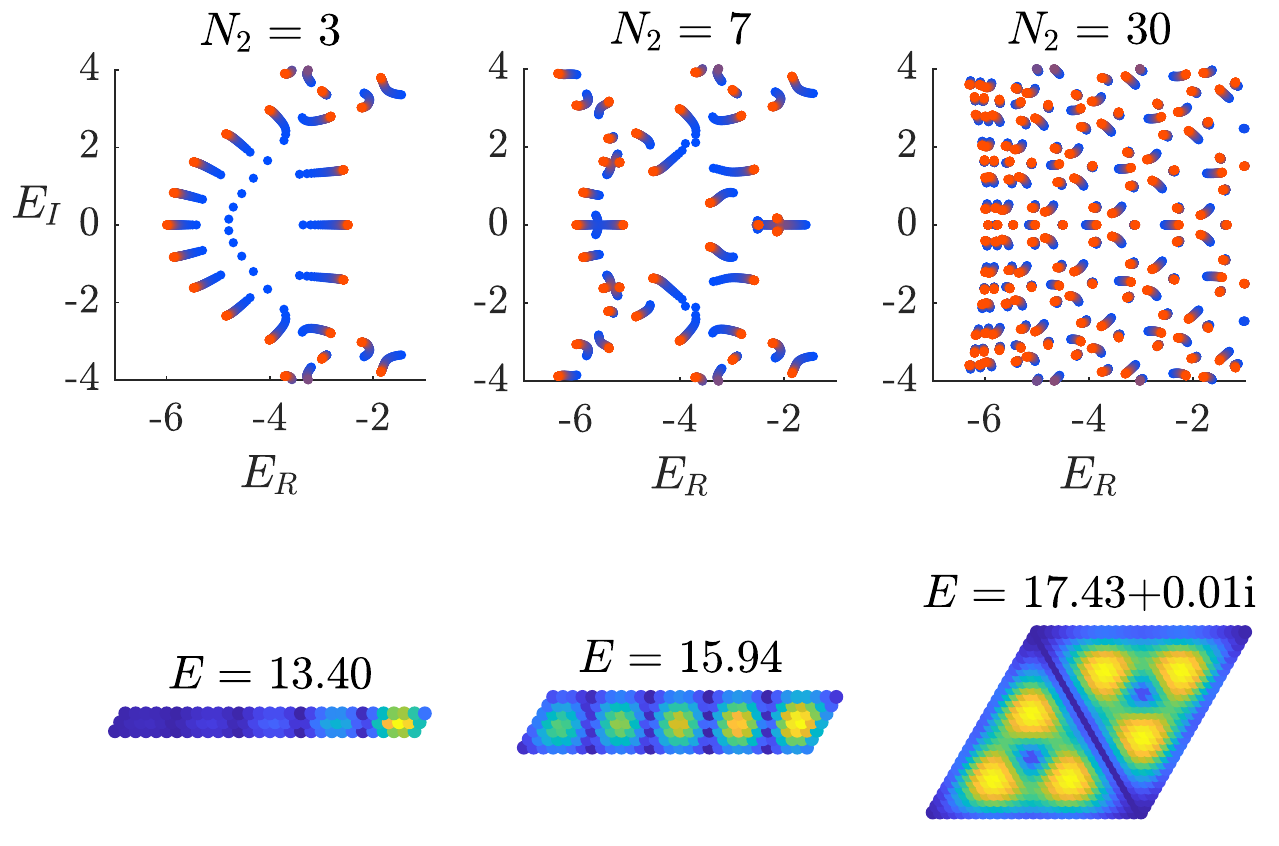}
	\caption{Upper panel: part of the spectrum of the triangular system for $t_l =1$, $t_r = 5$, $N_1 =30$ and different values of $N_2$.  The dots show how the eigenvalues change when we change $\delta_1$ in steps of $0.01$ from $0$ (blue dots) to $1$ (red dots). Lower panel: plot of the right expectation value of a representative state for each of the respective system sizes. We see the gradual disappearance of the skin effect as $N_2$ approaches $N_1$ when the right eigenstates become delocalized and the eigenvalues become less sensitive to boundary conditions.}
	\label{fig:triangle_eigenvalues}
\end{figure}

Up until now, we have effectively been studying a one-dimensional system with a large unit cell. We can, however, attempt to make some predictions of the behavior of the eigenvalues of the system with non-periodic boundary conditions in both directions. Namely, let us implement the balancing conditions from \textit{Case 3} in both directions of the lattice. Then, we end up with
\begin{equation}
	A(\delta_1) = \begin{pmatrix}
	0   & t_l   &        & \delta_1 t_r\\
	t_r &\ddots & \ddots &     \\
	    &\ddots & \ddots & t_l \\
	\delta_1 t_l & & t_r &0    
	\end{pmatrix},
\end{equation}
\begin{equation}
	B(\delta_1) = \begin{pmatrix}
	t_r &      & & \delta_1 t_l\\
	t_l &\ddots & &\\  
	    & \ddots &\ddots &\\
	    && t_l &t_r
	\end{pmatrix},
\end{equation}
and 
\begin{equation}
	C(\delta_1) = \begin{pmatrix}
	t_l & t_r &&\\
	& \ddots &\ddots &\\
	& &\ddots & t_r\\
	\delta_1 t_r && &t_l
	\end{pmatrix},
\end{equation}
which describes a triangular lattice, schematically shown in Fig.~\ref{fig:triangular_lattice}.
For this subclass of \textit{Case 3}, it turns out that if one implements the boundary conditions \textit{BC 1}, the eigenvalues are actually contained within the curve Eq.~\eqref{eq:parametrized_curve}.
One would expect that having implemented the balancing condition in both directions, the eigenvalues would show non-exponential sensitivity to changes in $\delta_1$ and $\delta_2$. Surprisingly enough, however, we find this to be the case only when $N_1 = N_2$. We numerically study the triangular lattice with open boundary conditions in one direction when going from open to periodic boundary conditions in the other. As expected, the eigenvalues still lie within the region bounded by the curve in Eq.~\eqref{eq:parametrized_curve} for all values of $N_1$ and $N_2$.
Moreover, the spectrum seems to have non-exponential sensitivity when $N_1 = N_2$.
However, when $N_1\neq N_2$, and in particular when $N_2\ll N_1$ or $N_1\ll N_2$, we start to see a higher sensitivity to boundary conditions and that the eigenstates start to localize at the end of the lattice. This can be seen in Fig.~\ref{fig:triangle_eigenvalues}, where we plot a part of the eigenvalue spectrum for a triangular lattice with $t_l = 1$ and $t_r = 5$ for different values of $N_2$ when keeping $\delta_2 = \delta'_2= 0$ and changing $\delta_1$ from $0$ to $1$ in steps of $0.01$. 
In this figure, we also plot the right expectation value $\langle \Pi_{\tilde\alpha,n} \rangle^{r,r} = | \psi^r_{\tilde\alpha,n}|^2$ for a representative eigenvector, for the different values of $N_2$. We see how the states show a significant localization to one of the sides of the system when $N_2\ll N_1$, but not when $N_2 = N_1$. We already encountered this behavior in Sec.~\ref{sec:unsolvable_system} for the Hamiltonian in Eq.\eqref{eq:small_triangle}, which describes the case $N_2 = 2$, where we saw that we do have a skin effect. Here, we thus see a gradual disappearance of the skin effect when we let $N_2$ approach $N_1$. We checked that the localization of the eigenstates is always `in the long direction of the system', so the localization due to the skin effect we observe here, is similar to the localization due to skin effect in the one-dimensional systems we studied earlier. 
To explain the presence of the skin effect in the triangular lattice model, we interpret the model as a one-dimensional model with a large unit cell of size $N_2$ and length $N_1$, such that $N_2 \ll N_1$.
In this way, we can use the phase winding formula \eqref{eq:winding}, which implicitly assumes the thermodynamic limit $N_1 \rightarrow \infty$.
To apply Eq.~\eqref{eq:winding}, we assume, for simplicity, that the system is open in the `2' direction. That means that the Bloch hamiltonian is a $N_2 \times N_2$ tridiagonal matrix
\begin{equation}
H(k) = \begin{pmatrix}
a & b & & 0\\
c & \ddots & \ddots &  \\
 & \ddots & \ddots & b  \\
0 & & c & a \\
\end{pmatrix} ,
\end{equation}
where
$a = t_l e^{-i k} + t_r e^{i k}$,
$b = t_r + t_l e^{i k}$,
$c = t_l + t_r e^{-i k}$.
We denote the determinant by $\det(N_2)$.
It is straightforward to show that
$\det(N_2) = a \det(N_2-1) - b c \det(N_2-2)$, while
$\det(1) = a$ and $\det(2) = a^2 - bc$.
Generically, $\det(N_2)$ will wind around some point in the complex plane,
implying that the system shows a skin effect, consistent with our results
that we for generic parameters observe a skin effect when $N_2 \neq N_1$.
The argument above does not imply that we should have a skin effect for all values $N_1$,
because we need the assumption that $N_1 \gg N_2$. Indeed, when $N_1$ is of the order of $N_2$, the system cannot be considered one-dimensional. Nevertheless, it is interesting to see that we do not observe a skin effect when $N_1 \sim N_2$ and both large, because in principle, there could have been.

Using the above approach, we can also try to find parameters for which the skin effect is absent, regardless of the system sizes. It turns out that by setting
$t_r / t_l = e^{i \phi}$, with $\phi$ real, the determinant takes
the form
$\det(N_2) = e^{i N_2 \phi/2} f(k,\phi)$,
where $f(k,\phi)$ is real and periodic in $k$.
So in this case, $\det(N_2)$ does not wind in the complex plane, and we do not have a skin effect. This is consistent with what we saw in Sec.~\ref{sec:unsolvable_system}.

The dependence of the skin effect on the ratio $N_1/N_2$ we observe here
seems to contradict the results in \cite{ZhYaFa2021}. Namely, in this reference, it is claimed that we have a skin effect in a two-dimensional system if and only if the spectral area is finite. As we have argued before, the spectral area of the triangular lattice is finite when we let the system size go to infinity, also when keeping the ratio $N_1/N_2\neq 1$ constant. This does also not seem to be taken into account by the exception of the geometry dependent skin effect, described in the same reference, where it is said that if the boundary of the system coincides with a mirror symmetry line of the lattice, the skin effect will disappear, even if the spectral area is finite. In our case, the boundaries of the system \emph{do} coincide with mirror symmetry lines, but we still have, for $N_1\neq N_2$, a localization of the eigenstates to one of the sides of the lattice, which implies that the mirror symmetry is not enough to prevent the skin effect from appearing.

We do not have analytical expressions for the eigenvalues in the case that $\delta_2 = \delta_2' = 0$, and $\delta_1$ interpolates between $0$ and $1$, so we cannot see analytically how the difference between $N_1$ and $N_2$ would affect the behavior of the eigenvalues in this case.
However, we can get a hint of this if we instead consider the system with the different boundary conditions from \textit{BC 2}. Then we get
\begin{equation}
    \tilde{H}_j = \begin{pmatrix}
	h_d  & h_l &        &        & \delta_1h_r\\
	h_r & h_d   & h_l     &        &   \\
	& h_r & \ddots & \ddots &   \\
	&    & \ddots & \ddots & h_l\\
	\delta_1h_l &    &        &  h_r     & h_d 
	\end{pmatrix},
\end{equation}
where
\begin{equation}
\begin{split}
    h_d &= t_r\omega_j\delta_2^{1/N_2}+t_l\omega_j^{-1}\delta_2^{-1/N_2},\\
    h_l &= t_l+t_r\omega_j^{-1}\delta_2^{-1/N_2},\\
    h_r &= t_r+t_l\omega_j\delta_2^{1/N_2}.
\end{split}
\end{equation}

We know the eigenvalues of this matrix. They are given by
\begin{equation}
\begin{split}
     \lambda_{jk} &=t_r\omega_j\delta_2^{1/N_2}+t_l\omega_j^{-1}\delta_2^{-1/N_2}\\
     &+2\sqrt{t_l+t_r\omega_j^{-1}\delta_2^{-1/N_2}}\sqrt{t_r+t_l\omega_j\delta_2^{1/N_2}}\cos(\tilde{\alpha}),
\end{split}
\end{equation}
where $\tilde{\alpha}$ is determined by the equation

\begin{equation}
\label{eq:alphatilde-triangular-bc2}
	\begin{split}
	&\delta_1\left(\left[\omega_j^{-1}\delta_2^{-1/N_2}\right]^{N_1/2}+\left[\omega_j\delta_2^{1/N_2}\right]^{N_1/2}\right)\\&-\frac{\sin[(N_1+1)\tilde\alpha]}{\sin(\tilde\alpha)}+\delta_1^2\frac{\sin[(N_1-1)\tilde\alpha]}{\sin(\tilde\alpha)} = 0,
	\end{split}
\end{equation} 
and we see that unless $N_1 = N_2$ there will be an exponential dependence on system size in $\tilde\alpha$.
To derive Eq.~\eqref{eq:alphatilde-triangular-bc2}, we assumed that $N_1$ is even, so that we do not have to worry about square roots. 
We note that this exponential dependence on the system size is somewhat different form the exponential dependence we encountered so far.
Here, it is the perturbing parameter $\delta_2$ that is raised to the power $N_1/N_2$, while earlier, it was some combination of the hopping parameters that was raised to the system size.

As we stated above, the boundary conditions we considered here are not the same as in Fig.~\ref{fig:triangle_eigenvalues}.
However, with the boundary conditions \textit{BC 2}, we do get similar behavior of the spectrum as in Fig.~\ref{fig:triangle_eigenvalues}.
Namely, depending on the ratio $N_1/N_2$, there is a skin effect (that is pronounced when $N_1 \ll N_2$). In the presence of the skin effect, the eigenvalues are also sensitive to changes in $\delta_2.$
Because of the similarity in behavior, we think that the mechanism at play in the case of Fig.~\ref{fig:triangle_eigenvalues}, is similar to the mechanism we just obtained analytically for the boundary conditions \textit{BC 2}.

We close the discussion on the triangular lattice system, by making a few remarks about the sensitivity of the eigenvalues when we do have a skin effect.
Because the exponential dependence on the system size in Eq.~\eqref{eq:alphatilde-triangular-bc2} has exponent $N_1/N_2$, one has to specify in which way one takes the large system size limit, when discussing the sensitivity of the eigenvalues as a function of system size.
If one fixes the ratio $N_1/N_2$, and take both $N_1$ and $N_2$ large, the sensitivity of the eigenvalues does not change in the large system size limit.
If, on the other hand, one fixes the size of one of the directions, the exponential sensitivity increases when making the other direction large.
In any case, if the eigenvalues are sensitive, this is due to the presence of the skin effect.
As stated above, the interesting aspects of the triangular lattice system are the presence of the skin effect, despite the fact that we have `balanced' parameters in both directions, and that the appearance of the skin effect depends on the ratio $N_1/N_2$.

\subsubsection{The Kagome lattice}

\begin{figure}
	\includegraphics[scale=0.12]{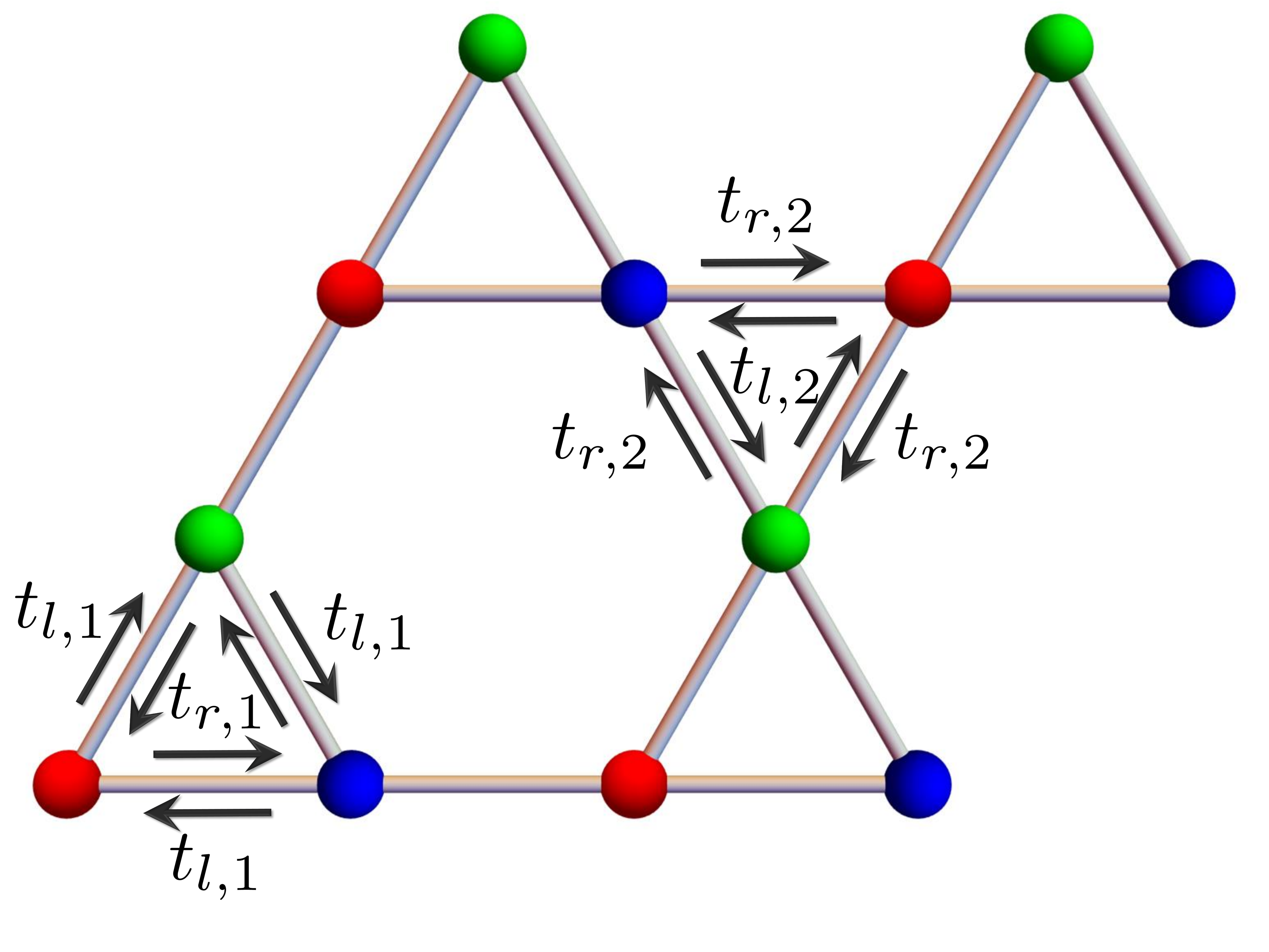}
	\caption{The Kagome lattice.}
	\label{fig:kagome_lattice}
\end{figure}

Perhaps a more interesting system, which, as is shown in \cite{EdKuBe2019}, supports boundary states, is the Kagome lattice. This system has similarities with the triangular lattice, and the results from the triangular case seem to be generalizable to the Kagome bulk states. Here we look at the Kagome lattice with hoppings according to Fig.~\ref{fig:kagome_lattice}. 

Also in this case, we observed numerically that for $N_1 = N_2$, the bulk spectrum has non-exponential sensitivity to the boundary conditions, while for $N_2 \ll N_1$, we seem to have a higher sensitivity among the bulk states. By studying the right eigenvectors of these insensitive eigenvalues, we notice that, just as in the case of the triangular lattice, for each eigenvector $\psi$, $\langle \Pi_n\rangle^{rr}$ becomes more localized the bigger the difference between $N_1$ and $N_2$ is, which implies that we get a skin effect when $N_1$ and $N_2$ deviate from each other. This phenomenon with the emerging skin effect shows that the explanation for the absence of the skin effect given in \cite{EdKuBe2019} is incomplete. There, the closed loops formed in the system were used as an explanation for not getting this build-up of the eigenstates in the system, but since the loops exist also in the case of $N_1\neq N_2$ there must be something more behind this.
We do currently not have an explanation for this behavior, but as we observed for the triangular lattice in the case of \emph{BC2}, the sensitivity depends on $N_1/N_2$, which is at least an indication that it is not unreasonable to expect a similar dependence on system size also in this case.

\subsection{Stacking SSH-chains}

\begin{figure}
	\includegraphics[scale=0.115]{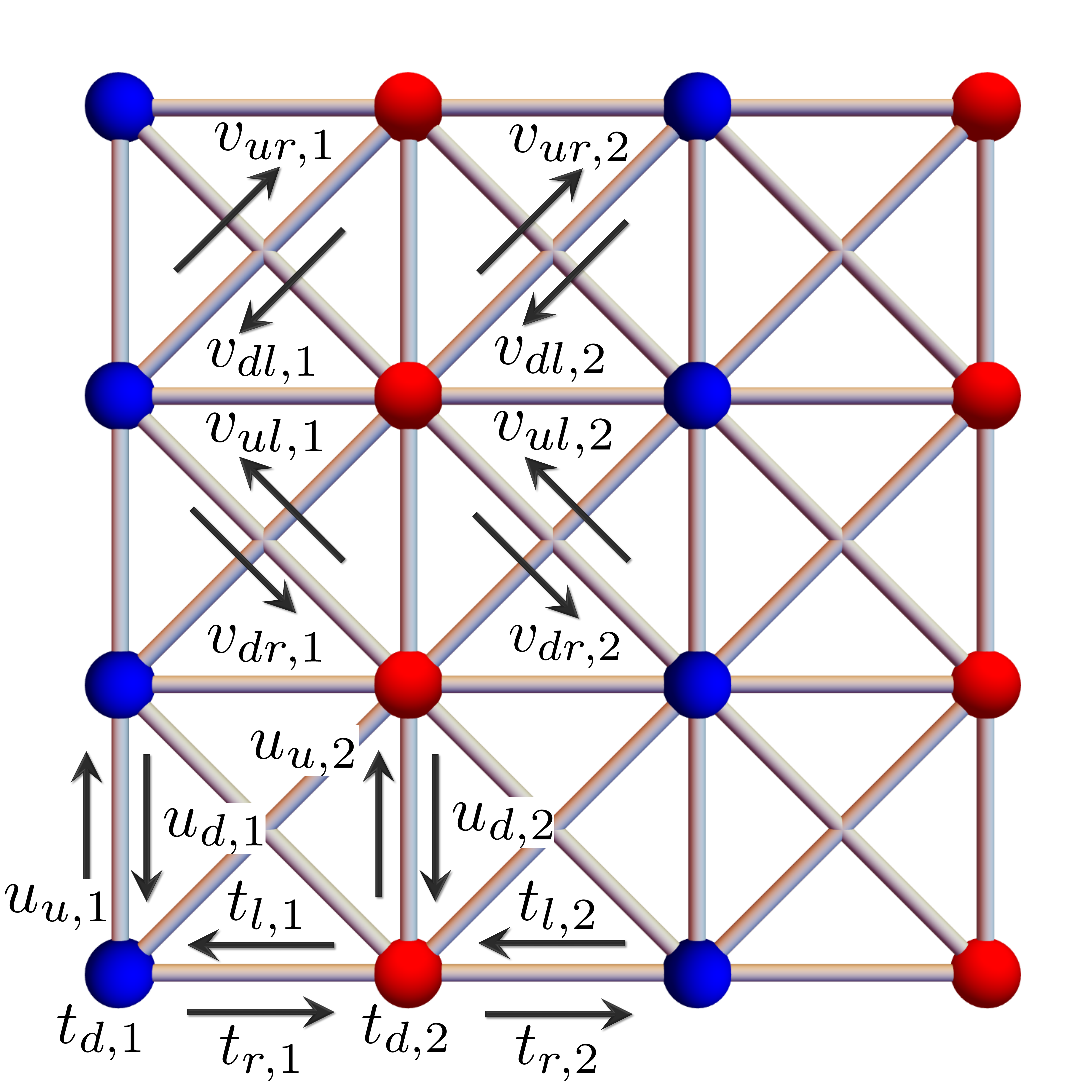}
	\caption{The lattice consisting of stacked SSH-chains.}
	\label{fig:ssh_like}
\end{figure}

Finally, we study the lattice in Fig.~\ref{fig:ssh_like}, which is obtained by stacking SSH-chains. 
In this case, we have (for $N_1$ even)
\begin{equation}
A(\delta_1) = \begin{pmatrix}
t_{d,1}  & t_{l,1} &        &        & \delta_1 t_{r,2}\\
t_{r,1} & t_{d,2}  & t_{l,2}     &        &   \\
& t_{r,2} & \ddots & \ddots &   \\
&    & \ddots & \ddots & t_{l,1}\\
\delta_1 t_{l,2} &    &        &  t_{r,1}    & t_{d,2}
\end{pmatrix},
\end{equation}

\begin{equation}
B(\delta_1) = \begin{pmatrix}
u_{d,1}  & v_{dl,1} &        &        & \delta_1 v_{dr,2}\\
v_{dr,1} & u_{d,2}  & v_{dl,2}     &        &   \\
& v_{dr,2} & \ddots & \ddots &   \\
&    & \ddots & \ddots & v_{dl,1}\\
\delta_1 v_{dl,2} &    &        &  v_{dr,1}    & u_{d,2}
\end{pmatrix},
\end{equation}

\begin{equation}
C(\delta_1) = \begin{pmatrix}
u_{u,1}  & v_{ul,1} &        &        & \delta_1 v_{ur,2}\\
v_{ur,1} & u_{u,2}  & v_{ul,2}     &        &   \\
& v_{ur,2} & \ddots & \ddots &   \\
&    & \ddots & \ddots & v_{ul,1}\\
\delta_1 v_{ul,2} &    &        &  v_{ur,1}    & u_{u,2}
\end{pmatrix},
\end{equation}

which means that we get
\begin{equation}
\tilde{H}_j = \begin{pmatrix}
h_{d,1}  & h_{l,1} &        &        & \delta_1t_{r,2}\\
h_{r,1} & h_{d,2}  & h_{l,2}     &        &   \\
& h_{r,2} & \ddots & \ddots &   \\
&    & \ddots & \ddots & h_{l,1}\\
\delta_1h_{l,2} &    &        &  h_{r,1}    & h_{d,2}
\end{pmatrix},
\end{equation}
where 
\begin{equation}
\begin{split}
h_{d,1} &= t_{d,1}+\omega_ju_{d,1}+\omega_j^{-1}u_{u,1},\\
h_{d,2} &= t_{d,2}+\omega_ju_{d,2}+\omega_j^{-1}u_{u,2},\\
h_{l,1} &= t_{l,1}+\omega_jv_{dl,1}+\omega_j^{-1}v_{ul,1},\\
h_{l,2} &= t_{l,2}+\omega_jv_{dl,2}+\omega_j^{-1}v_{ul,2},\\
h_{r,1} &= t_{r,1}+\omega_jv_{dr,1}+\omega_j^{-1}v_{ur,1},\\
h_{r,2} &= t_{r,2}+\omega_jv_{dr,2}+\omega_j^{-1}v_{ur,2},
\end{split}
\end{equation}
and $\omega_j = e^{2 \pi i j/N_2}$. From Eqs.~\eqref{eq:ssh_eigenvalues_potential} and \eqref{eq:ssh_alphaeq} in section \ref{sec:ssh}, we know how to find the eigenvalues of this matrix. Namely, they are given by
\begin{equation}
\begin{split}
	&\lambda_{j\tilde{\alpha}} = \frac{h_{d,1}+h_{d,2}}{2}\pm\Biggl[\left(\frac{h_{d,1}-h_{d,2}}{2}\right)^2+h_{r,1}h_{l,1}\\&+h_{r,2}h_{l,2}+2\cos(\tilde{\alpha})\sqrt{h_{r,1}}\sqrt{h_{l,1}}\sqrt{h_{r,2}}\sqrt{h_{l,2}}\Biggr]^{1/2},
\end{split}
\end{equation}
where $\tilde{\alpha}$ is determined by the equation
\begin{align}
&0 = -\frac{\sin[\tilde{\alpha}(N_1/2+1)]}{\sin(\tilde{\alpha})} + \delta_1^2 \frac{\sin[\tilde{\alpha}(N_1/2-1)]}{\sin(\tilde{\alpha})}
 \nonumber\\
&+(\delta_1^2 - 1)
\frac{
\sqrt{h_{l,2}}\sqrt{h_{r,2}}
}
{
\sqrt{h_{l,1}}\sqrt{h_{r,1}}
}
\frac{\sin(\tilde{\alpha}N_1/2)}{\sin(\tilde{\alpha})} \nonumber\\
&
+\delta_1
\left[
\left(\frac{\sqrt{h_{r,1}}\sqrt{h_{r,2}}}{\sqrt{h_{l,1}}\sqrt{h_{l,2}}}\right)^{N_1/2}+
\left(\frac{\sqrt{h_{l,1}}\sqrt{h_{l,2}}}{\sqrt{h_{r,1}}\sqrt{h_{r,2}}}\right)^{N_1/2}
\right] \ .
\end{align}

\begin{figure*}[t]
	\includegraphics[width=\linewidth]{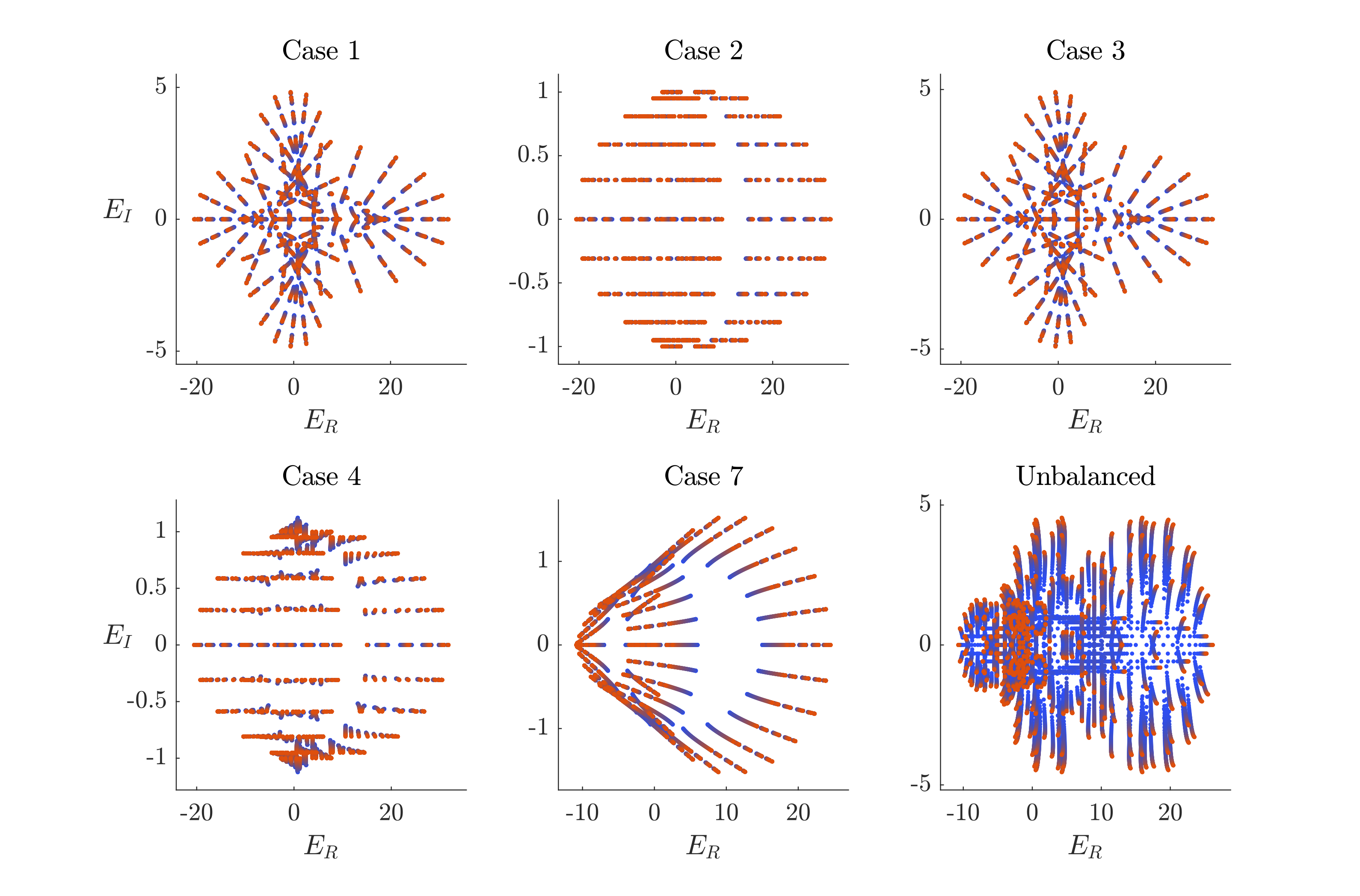}
	\caption{Plots of the eigenvalues of different lattices of stacked SSH-chains. The dots show how the eigenvalues change when we change $\delta_1$ in steps of $0.01$ from $0$ (blue dots) to $1$ (red dots). In all cases, the system size is $N_1 = N_2 = 30$, with $\delta_2 = \delta_2' = 0$, and the other parameters in each case can be found in the main text.}
	\label{fig:stacked_ssh_cases}
\end{figure*}

As before, we note that if
\begin{equation}\label{eq:stacked_ssh_condition}
	\left|\frac{\sqrt{h_{r,1}}\sqrt{h_{r,2}}}{\sqrt{h_{l,1}}\sqrt{h_{l,2}}}\right| = 1,
\end{equation}
the exponential sensitivity of the spectrum disappears. Similar to the results in the previous section, we see that this gives rise to several different cases, and we will now list some of them together with plots of the eigenvalues of some representative systems. Assuming that all parameters are real, we first list four cases where individual hoppings are related to each other:

\vspace{0.5cm}
\textit{Case 1.} The case $h_{r,1} = h_{l,1}$ and $h_{r,2} = h_{l,2}$. This implies
	\begin{equation}
		\begin{split}
		t_{r,1} &= t_{l,1}, v_{dr,1} = v_{dl,1}, v_{ur,1} = v_{ul,1},\\
		t_{r,2} &= t_{l,2}, v_{dr,2} = v_{dl,2}, v_{ur,2} = v_{ul,2}.
		\end{split}
	\end{equation}

In Fig.\ref{fig:stacked_ssh_cases}, we plot the eigenvalues of a system with $t_{r,1} = t_{l,1} = 1$, $v_{dr,1} = v_{dl,1} = 3$, $v_{ur,1} = v_{ul,1} = 5$, $t_{r,2} = t_{l,2} = 2$,$v_{dr,2} = v_{dl,2} = 4$, $v_{ur,2} = v_{ul,2} = 6$, $t_{d,1} = 1$, $t_{d,2} = 4$, $u_{u,1} = 2$, $u_{u,2} = 5$, $u_{d,1} = 3$ and $u_{d,2} = 6$ for $N_1 = N_2 = 20$ when we change $\delta_1$ in steps of $0.01$ from $\delta_1 = 0$ (blue dots) to $\delta_1 = 1$ (red dots).

\vspace{0.5cm}	
\textit{Case 2.}  The case $h_{r,1} = h_{l,1}^*$ and $h_{r,2} = h_{l,2}^*$. This implies
	\begin{equation}
	\begin{split}
	t_{r,1} &= t_{l,1}, v_{dr,1} = v_{ul,1}, v_{ur,1} = v_{dl,1},\\
	t_{r,2} &= t_{l,2}, v_{dr,2} = v_{ul,2}, v_{ur,2} = v_{dl,2}.
	\end{split}
	\end{equation}
	
In Fig.\ref{fig:stacked_ssh_cases}, we plot the eigenvalues of a system with $t_{r,1} = t_{l,1} = 1$, $v_{dr,1} = v_{ul,1} = 3$, $v_{ur,1} = v_{dl,1} = 5$, $t_{r,2} = t_{l,2} = 2$,$v_{dr,2} = v_{ul,2} = 4$, $v_{ur,2} = v_{dl,2} = 6$, $t_{d,1} = 1$, $t_{d,2} = 4$, $u_{u,1} = 2$, $u_{u,2} = 5$, $u_{d,1} = 3$ and $u_{d,2} = 6$ for $N_1 = N_2 = 20$ when we change $\delta_1$ in steps of $0.01$ from $\delta_1 = 0$ (blue dots) to $\delta_1 = 1$ (red dots).	

\vspace{0.5cm}	
\textit{Case 3.}  The case $h_{r,1} = h_{l,2}$ and $h_{r,2} = h_{l,1}$. This implies
	\begin{equation}
	\begin{split}
	t_{r,1} &= t_{l,2}, v_{dr,1} = v_{dl,2}, v_{ur,1} = v_{ul,2},\\
	t_{r,2} &= t_{l,1}, v_{dr,2} = v_{dl,1}, v_{ur,2} = v_{ul,1}.
	\end{split}
	\end{equation}
	
In Fig.\ref{fig:stacked_ssh_cases}, we plot the eigenvalues of a system with $t_{r,1} = t_{l,2} = 1$, $v_{dr,1} = v_{dl,2} = 3$, $v_{ur,1} = v_{ul,2} = 5$, $t_{r,2} = t_{l,1} = 2$,$v_{dr,2} = v_{dl,1} = 4$, $v_{ur,2} = v_{ul,1} = 6$, $t_{d,1} = 1$, $t_{d,2} = 4$, $u_{u,1} = 2$, $u_{u,2} = 5$, $u_{d,1} = 3$ and $u_{d,2} = 6$ for $N_1 = N_2 = 20$ when we change $\delta_1$ in steps of $0.01$ from $\delta_1 = 0$ (blue dots) to $\delta_1 = 1$ (red dots).	

\vspace{0.5cm}	
\textit{Case 4.} The case $h_{r,1} = h_{l,2}^*$ and $h_{r,2} = h_{l,1}^*$. This implies
	\begin{equation}
	\begin{split}
	t_{r,1} &= t_{l,2}, v_{dr,1} = v_{ul,2}, v_{ur,1} = v_{dl,2},\\
	t_{r,2} &= t_{l,1}, v_{dr,2} = v_{ul,1}, v_{ur,2} = v_{dl,1}.
	\end{split}
	\end{equation}
	
In Fig.\ref{fig:stacked_ssh_cases}, we plot the eigenvalues of a system with $t_{r,1} = t_{l,2} = 1$, $v_{dr,1} = v_{ul,2} = 3$, $v_{ur,1} = v_{dl,2} = 5$, $t_{r,2} = t_{l,1} = 2$,$v_{dr,2} = v_{ul,1} = 4$, $v_{ur,2} = v_{dl,1} = 6$, $t_{d,1} = 1$, $t_{d,2} = 4$, $u_{u,1} = 2$, $u_{u,2} = 5$, $u_{d,1} = 3$ and $u_{d,2} = 6$ for $N_1 = N_2 = 20$ when we change $\delta_1$ in steps of $0.01$ from $\delta_1 = 0$ (blue dots) to $\delta_1 = 1$ (red dots).	

\vspace{0.5cm}
We notice that all these cases imply an overall balancing of the hoppings in the unit cell. In general, however, Eq.~\eqref{eq:stacked_ssh_condition} implies
\begin{equation}
	h_{r,1}h_{r,2} = e^{2\pi ir/N_2}h_{l,1}h_{l,2},
\end{equation}
where $r\in\mathbb{R}$, which means that we also can get conditions consisting of relations between products of hopping parameters. We list a few such cases.

\vspace{0.5cm}
\textit{Case 5.} For $r = 0$, we get
	\begin{equation}
		\begin{cases}
		\begin{aligned}
		t_{r,1}t_{r,2}&+v_{dr,1}v_{ur,2}+v_{ur,1}v_{dr,2}\\ &= t_{l,1}t_{l,2}+v_{dl,1}v_{ul,2}+v_{ul,1}v_{dl,2}, 
		\end{aligned}\\
		t_{r,1}v_{dr,2}+v_{dr,1}t_{r,2} = t_{l,1}v_{dl,2}+v_{dl,1}t_{l,2},\\
		t_{r,1}v_{ur,2}+v_{ur,1}t_{r,2} = t_{l,1}v_{ul,2}+v_{ul,1}t_{l,2},\\
		v_{dr,1}v_{dr,2} = v_{dl,1}v_{dl,2},\\
		v_{ur,1}v_{ur,2} = v_{ul,1}v_{ul,2}.
		\end{cases}
	\end{equation}

\vspace{0.5cm}	
\textit{Case 6.}  For $r = j$, we get
	\begin{equation}
	\begin{cases}
	\begin{aligned}
	t_{r,1}t_{r,2}&+v_{dr,1}v_{ur,2}+v_{ur,1}v_{dr,2}\\ &= t_{l,1}v_{ul,2}+v_{ul,1}t_{l,2}, 
	\end{aligned}\\
	\begin{aligned}
	t_{r,1}v_{dr,2}&+v_{dr,1}t_{r,2}\\ &= t_{l,1}t_{l,2}+v_{dl,1}v_{ul,2}+v_{ul,1}v_{dl,2},
	\end{aligned}\\
	t_{r,1}v_{ur,2}+v_{ur,1}t_{r,2} = v_{ul,1}v_{ul,2},\\
	v_{dr,1}v_{dr,2} =  t_{l,1}v_{dl,2}+v_{dl,1}t_{l,2},\\
	v_{ur,1}v_{ur,2} = 0,\\
	v_{dl,1}v_{dl,2} = 0.
	\end{cases}
	\end{equation}

\vspace{0.5cm}
\textit{Case 7.}  For $r = 2j$, we get
	\begin{equation}
	\begin{cases}
	t_{r,1}t_{r,2}+v_{dr,1}v_{ur,2}+v_{ur,1}v_{dr,2}=v_{ul,1}v_{ul,2}\\ 
	t_{r,1}v_{dr,2}+v_{dr,1}t_{r,2} = t_{l,1}v_{ul,2}+v_{ul,1}t_{l,2},\\
	t_{r,1}v_{ur,2}+v_{ur,1}t_{r,2} = 0,\\
	v_{dr,1}v_{dr,2} = t_{l,1}t_{l,2}+v_{dl,1}v_{ul,2}+v_{ul,1}v_{dl,2},\\
	v_{ur,1}v_{ur,2} = 0,\\
	t_{l,1}v_{dl,2}+v_{dl,1}t_{l,2} = 0,\\
	v_{dl,1}v_{dl,2} = 0.
	\end{cases}
	\end{equation}

In Fig.~\ref{fig:stacked_ssh_cases}, we plot the eigenvalues of a system with $t_{r,1} = t_{l,1} = 1$, $v_{dl,1} = v_{dl,2} = v_{ur,1} = v_{ur,2} = 0$, $t_{r,2} = 6$, $t_{l,2} = 8$,$v_{dr,1} =8/3$, $v_{ul,1} = 2$, $v_{ul,2} = 3$, $t_{d,1} = 1$, $t_{d,2} = 4$, $u_{u,1} = 2$, $u_{u,2} = 5$, $u_{d,1} = 3$ and $u_{d,2} = 6$ for $N_1 = N_2 = 20$ when we change $\delta_1$ in steps of $0.01$ from $\delta_1 = 0$ (blue dots) to $\delta_1 = 1$ (red dots).

\vspace{0.5cm}
While the cases 1-4 give us equalities between individual hopping parameters, we see that cases 5-7 give us relations between products of adjacent hopping parameters. We also note that in the cases 5-7 we still need some kind of balance in the left-right direction, even though it is not as straightforward as in previous cases.
It is interesting to see that by stacking SSH chains, there are indeed more complicated ways in which the system can be balanced, because of the fact that one has more parameters to play with. We note, just as in the case with stacked HN-chains, that the hopping parameters in the direction of the periodic boundary conditions do not affect the sensitivity of the spectrum at all. 

For comparison, we also plot the eigenvalues of an unbalanced system with parameters
$t_{d,1} = 1$,
$u_{u,1} = 2$,
$u_{d,1}= 3$,
$t_{d,2} = 4$,
$u_{u,2} = 5$,
$u_{d,2}= 6$,
$t_{r,1} = 1$,
$t_{r,2} = 2$,
$t_{l,1} = 3$,
$t_{l,2} = 4$,
$v_{dr,1} = 1$,
$v_{dr,2} = 2$,
$v_{dl,1} = 3$,
$v_{dl,2} = 4$,
$v_{ur,1} = 1$,
$v_{ur,2} = 2$,
$v_{ul,1} = 3$
and
$v_{ul,2} = 4$
for $N_1 = N_2 = 20$ when we change $\delta_1$ in steps of $0.01$ from $\delta_1 = 0$ (blue dots) to $\delta_1 = 1$ (red dots).

From the plots in Fig.~\ref{fig:stacked_ssh_cases}, we see a significant difference in the behavior of the spectrum between the balanced and a representative unbalanced case when we change the boundary conditions.

We would like to comment on \textit{Case 4} in Fig.~\ref{fig:stacked_ssh_cases}. This case does not look `as neat' at the \textit{Cases 1-3}.
One could wonder if this is due to numerical inaccuracy, despite the fact that this is a balanced case. We therefore checked the numerical results against our analytical results, and found perfect agreement between the two (that is, agreement up to machine precision).

\section{Discussion}
\label{sec:discussion}

In this paper, we obtain (almost) analytical expressions for the eigenvalues of non-Hermitian one-dimensional one-band models with arbitrary boundary conditions. We use the analytical expressions to analyze the sensitivity of the eigenvalues to the boundary conditions in several one-dimensional systems. Using the one-dimensional models, we construct two-dimensional models with arbitrary boundary conditions in one direction, and either periodic or a particular deformation of periodic boundary conditions in the other direction. 

We find that for most parameter values, we have eigenvalues that are exponentially sensitive to boundary conditions, but that there are parameter values for which this exponential sensitivity disappears. In this case, the behavior of the eigenvalues, apart from the fact that the eigenvalues can still be complex, is similar to what we expect in a Hermitian system.

In one-dimensional chains with nearest neighbor hoppings, we find the rather intuitive result that in order for the spectrum to behave similar to the Hermitian case, the hoppings to the left should balance those to the right. This seems to be a necessary, but not sufficient condition for 'Hermitian behavior', which is a bit surprising. Even more surprising is the fact that in the case of longer range hoppings it might not be possible at all to balance the system. This indicates that the intuitive understanding of why the system shows 'Hermitian behavior' in some cases is lacking, and it would be interesting to understand why this is. 

In two dimensions, the situation is more complicated. Ideally, one would analytically like to study arbitrary boundary conditions in both directions, but this turns out to be out of reach. One can, however, study the case with open boundary conditions in one direction, and a deformation of periodic boundary conditions in the other. One can then use these results to explain the numerically obtained behavior of two-dimensional systems with open boundaries in both directions.

In particular, we studied a rather general two-dimensional model obtained by stacking HN-chains. This model contains a hopping model on the triangular lattice as a special case. By studying this model, with periodic boundary conditions in one direction, and arbitrary boundary conditions in the other, we obtain a condition on the hopping parameters, such that the system is not exponentially sensitive. This condition amounts to a local balancing of the hopping in the direction with arbitrary boundary conditions, as in the one-dimensional case. If one implements this balancing condition in both directions, one could expect that the system is not exponentially sensitive. Our numerical results show that this is however not the case. Only if the system size is the same in both directions, the spectrum has non-exponential sensitivity. We obtained similar results for the Kagome lattice.

Clearly, obtaining a full understanding of the behavior and sensitivity of the eigenvalues of two-dimensional systems with open or arbitrary boundary conditions in both directions is a question of great interest for future investigation. So far, in contrast to the one-dimensional case where the winding number can be used, there is no general condition for the existence of the skin effect in two-dimensional systems, even though there have been attempts to come up with such conditions \cite{ZhYaFa2021}, but these conditions fail to capture the dependence on the aspect ratio of the system. In addition, our results also indicate that the picture put forward in \cite{EdKuBe2019}, namely that "loops" in the system would prevent the skin effect, is incomplete. 

Apart from stacking HN-chains to obtain two-dimensional models, we also studied two-dimensional models obtained by stacking non-Hermitian SSH-chains. In this case, the balancing conditions become more complicated, in comparison to the case where we stacked HN-chains. Namely, the balancing condition for the SSH case, becomes an equation for sums of product of hopping parameters. Despite the more complicated condition, one can still find non-trivial solutions, and indeed cases for which the system is balanced, thus without the exponential sensitivity to the boundary conditions.

In the current paper, we considered the eigenvalue stability of non-Hermitian hopping models. It would certainly be interesting to extend the current analysis to models including pairing terms. To deal with pairing terms, one could use the method of Lieb, Schultz and Mattis \cite{LiScMa1961}, and apply it to the non-Hermitian case. One can speculate that it should be possible to find balanced systems, even in the presence of pairing terms.

\section*{Acknowledgements}
We thank Emil J. Bergholtz, Vatsal Dwivedi, Jonas Larson, Iman Mahyeah, Eva Mossberg and Kang Yang for useful discussions. 
The research of E.E. was funded in part by the Swedish Research Council (VR) and the Knut and Alice Wallenberg Foundation. The research of E.A. was sponsored in part by the Swedish Research Council (VR).

\appendix

\section{Some details for the NH model}
\label{app:nh-details}

In this appendix, we give some more details of the non-Hermitian hopping model we consider in Sec.~\ref{sec:hn-model}.
Namely, we show that $\tilde\alpha$ is real under certain conditions, and we provide the solution for a slightly more general model.

\subsection{Proof that $\tilde{\alpha}$ is real for the HN model under certain conditions}
\label{app:alpha_real}

In this section, we show that Eq.~\eqref{eq:hn-alpha-tilde}, repeated here for convenience,
\begin{equation}
\label{eq-app:hn-alpha-tilde}
	\sin([N+1]\tilde{\alpha})-\delta^2\sin([N-1]\tilde{\alpha})= 2\delta\cos(\theta)\sin(\tilde{\alpha}) \ ,
\end{equation}
only has real solutions for $\tilde{\alpha}$, provided that $-1 \leq \delta \leq 1$ and $\theta \in \mathbb{R}$, that is, for $|t_l| = |t_r|$.
Under these conditions, the eigenvalues of the Hatano-Nelson model lie on a line-segment in the complex plane.
We recall that the trivial solutions $\tilde\alpha = 0, \pi$ should be discarded (unless there are multiple solutions at these values), because they do not lead to valid eigenstates.
From the argument below, it will become clear that we can assume that $\cos(\theta) = \pm 1$, because $| \cos (\theta) | < 1$ follows easily. For concreteness, we assume that $\cos(\theta) = 1$.

To show that Eq.~\eqref{eq-app:hn-alpha-tilde} (with $\cos(\theta) = 1$) only has real solutions for $-1 \leq \delta \leq 1$, we start by noticing that for $\delta = \pm 1$, the equation simplifies to
\begin{equation*}
	 2 \cos(N\tilde{\alpha})\sin(\tilde{\alpha}) = \pm 2\sin(\tilde{\alpha}) \ .
\end{equation*}
We find that the independent solutions for $\tilde\alpha$ are indeed real, and given by
$\tilde\alpha =  p \pi/N$, with $p$ even for $\delta = 1$ and $p$ odd for $\delta = -1$.
Discarding the solutions coming from $\sin(\tilde\alpha)=0$, we find that all solutions are double solutions.
To pick an independent set of solutions, we take the double solutions with $0 < p < N$, but only one solution for $p=0$ and $p=N$.

\begin{figure}[t]
\includegraphics[width=\columnwidth]{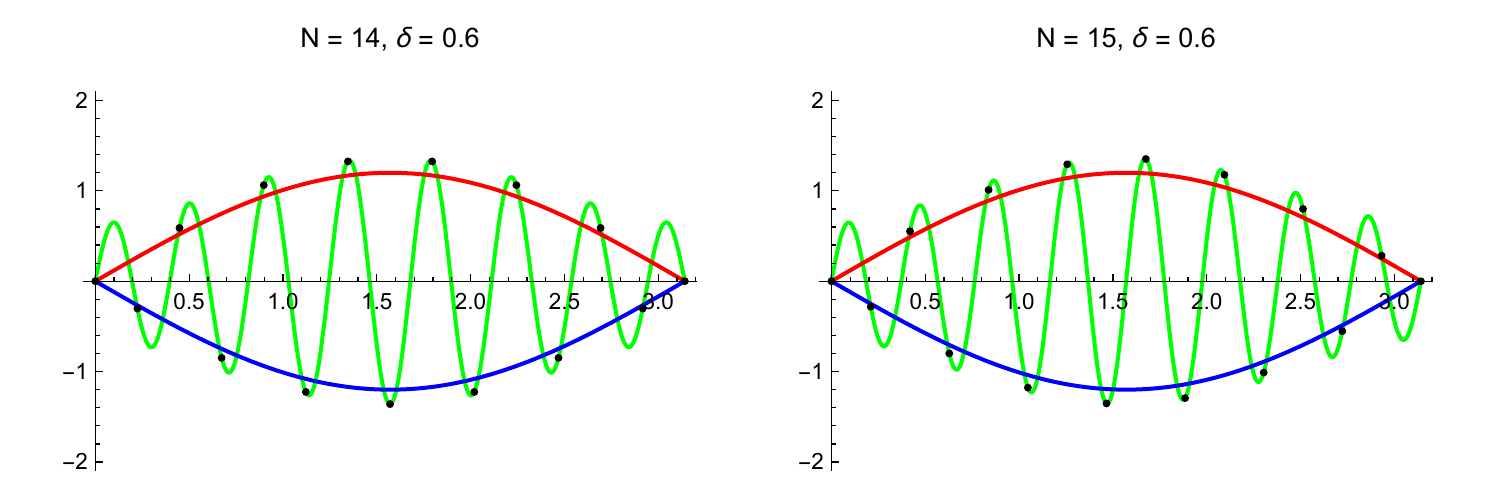}
\caption{Plot of the LHS of Eq.~\eqref{eq-app:hn-alpha-tilde}, as well as the RHS (with $\cos(\theta) = +1,-1$), as a function of $\tilde{\alpha}$ for $N=14,15$ and $\delta = 0.6$, in green, red and blue, respectively.
The black dots indicate the LHS, at the values $\tilde{\alpha} = p \pi/N$, for integer $p$.}
\label{fig:alpha-tilde-plot}
\end{figure}

The idea behind the proof is to show that upon decreasing $\delta$ from $\delta = 1$, one does not loose any of these real solutions.
To this end, we will show below that
$$
\sin([N+1]\tilde{\alpha})-\delta^2\sin([N-1]\tilde{\alpha}) > 2\delta\sin(\tilde{\alpha}) 
$$
for $\tilde{\alpha} = p \pi/N$ with $p$ even and
$$
\sin([N+1]\tilde{\alpha})-\delta^2\sin([N-1]\tilde{\alpha}) <  -2\delta\sin(\tilde{\alpha}) 
$$
for $\tilde{\alpha} = p \pi/N$ with $p$ odd.
In other words, $\sin([N+1]\tilde{\alpha})-\delta^2\sin([N-1]\tilde{\alpha})$ is alternatingly larger than
$2\delta \sin(\tilde{\alpha})$ and smaller than $-2\delta \sin(\tilde{\alpha})$.
We plot $\sin([N+1]\tilde{\alpha})-\delta^2\sin([N-1]\tilde{\alpha})$ and $\pm 2 \delta \sin(\tilde{\alpha})$ in Fig.~\ref{fig:alpha-tilde-plot},
together with the points $\tilde{\alpha} = p \pi/N$, to show this behavior.
This implies that the double real solutions for $\tilde{\alpha} = p \pi/N$ with $1\leq p \leq N-1$ for $\delta = 1$,
become distinct real solutions for $0\leq \delta < 1$.
It remains to be shown that the solutions $\tilde{\alpha} = 0,\pi$ for $\delta = 1$, remain real upon decreasing $\delta$ to zero.
To do this, we take the derivative of both sides of Eq.~\eqref{eq-app:hn-alpha-tilde} at $\tilde{\alpha}=0$; a simple calculation shows that the derivative at $\tilde{\alpha}=0$ of the LHS is larger than the derivative of the RHS for $0\leq \delta < 1$ (see also Fig.~\ref{fig:alpha-tilde-plot}).
It follows that the solution $\tilde{\alpha} = 0$ for $\delta = 1$ remains real for $0\leq \delta < 1$.
That this also is true for the solution $\tilde{\alpha} = \pi$ follows similarly.

To complete the argument, we need to show that 
$\sin([N+1]\tilde{\alpha})-\delta^2\sin([N-1]\tilde{\alpha}) - 2\delta\sin(\tilde{\alpha}) > 0$
for $\tilde{\alpha} = p \pi/N$ with $p$ even and $0\leq\delta < 1$.
Evaluating the LHS, one finds $(d-1)^2 \sin(p \pi/N)$, giving the required result.
Similarly, we find that 
$\sin([N+1]\tilde{\alpha})-\delta^2\sin([N-1]\tilde{\alpha}) + 2\delta\sin(\tilde{\alpha}) < 0$
for $\tilde{\alpha} = p \pi/N$ with $p$ odd and $0\leq\delta < 1$.

The proof works similarly when $-1 \leq \cos(\theta) < 1$, as well for $-1\leq \delta < 0$, which means we covered all cases, and we are done.

\subsection{Solution for a slightly more general hopping model}
\label{app:hn-more-general}

In this section, we provide the solution for a slightly more general non-Hermitian hopping model.
In particular, we consider the following hamiltonian (here given for $N=5$ sites)
\begin{equation}
\label{eq:mat-nhh}
H = \begin{pmatrix}
\epsilon_1 & t_l & 0 & 0 & \delta_r t_r\\
t_r & 0 & t_l & 0 & 0\\
0 & t_r & 0 & t_l & 0\\
0 & 0 & t_r & 0 & t_l\\
\delta_l t_l & 0 & 0 & t_r & \epsilon_N\\
\end{pmatrix} \ .
\end{equation}
The method to obtain the eigenvalues is identical to the one used in the main text in Sec.~\ref{sec:hn-model}, so we simply state the result here.
The eigenvalues of this modified non-Hermitian hopping model are given by
\begin{equation}
\label{eq:ev-nhh}
\lambda_{\tilde\alpha} = 2 \sqrt{t_r} \sqrt{t_l} \cos(\tilde\alpha) \ ,
\end{equation}
where $\tilde\alpha$ satisfies the transcendental equation
\begin{align}
\label{eq:alpha-tilde-nhh}
&- \frac{\sin((N+1)\tilde\alpha)}{\sin\tilde\alpha}
+ \frac{(\epsilon_1+\epsilon_N)}{\sqrt{t_r}\sqrt{t_l}} \frac{\sin(N \tilde\alpha)}{\sin\tilde\alpha}
\\
&+ (\delta_r \delta_l- \frac{\epsilon_1 \epsilon_N}{t_l t_r}) \frac{\sin((N-1)\tilde\alpha)}{\sin\tilde\alpha}
+\delta_l \frac{t_l^{N/2}}{t_r^{N/2}}
+\delta_r \frac{t_r^{N/2}}{t_l^{N/2}}
= 0 \ .
\nonumber
\end{align}

For $\delta_l = \delta_r = 0$, this case was studied in \cite{Lo1992}.

\section{Some details on the SSH chain}
\label{app:ssh}
In this appendix, we give some more details on the non-Hermitian SSH model.

\subsection{Solution for even length chains}
\label{app:ssh-even-length}
For completeness, we state the solution for the eigenvalues of the non-Hermitian SSH chain, Eq.~\eqref{eq:ssh_hamiltonian}, but in the presence of the perturbations $\delta_r$ and $\delta_l$ in the upper right and lower left corner respectively (i.e., with matrix elements $\delta_r t_{r,2}$ and $\delta_l t_{l,2}$ respectively).

The eigenvalues are as before,
\begin{equation}
\label{eq:eivenvalue_app}
    \lambda_{\tilde{\alpha}} = 
    \sqrt{t_{l,1}t_{r,1} + t_{l,2}t_{r,2} + 2 \cos(\tilde\alpha) \sqrt{t_{l,1}}\sqrt{t_{r,1}}\sqrt{t_{l,2}}\sqrt{t_{l,2}}} \ ,
\end{equation}
where $\tilde\alpha$ now is a solution of the equation
\begin{align}
&-\frac{\sin(\tilde{\alpha}(N/2+1))}{\sin(\tilde{\alpha})} +\delta_l \delta_r \frac{\sin(\tilde{\alpha}(N/2-1))}{\sin(\tilde{\alpha})}
 \nonumber\\
&+ (\delta_l\delta_r - 1)
\frac{
\sqrt{t_{l,2}}\sqrt{t_{r,2}}
}
{
\sqrt{t_{l,1}}\sqrt{t_{r,1}}
}
\frac{\sin(\tilde{\alpha}N/2)}{\sin(\tilde{\alpha})}\nonumber\\
&+\left[
\delta_r
\left(\frac{\sqrt{t_{r,1}}\sqrt{t_{r,2}}}{\sqrt{t_{l,1}}\sqrt{t_{l,2}}}\right)^{N/2}+
\delta_l
\left(\frac{\sqrt{t_{l,1}}\sqrt{t_{l,2}}}{\sqrt{t_{r,1}}\sqrt{t_{r,2}}}\right)^{N/2}
\right] = 0 \ .
\end{align}

\subsection{Solution for odd length chains}
\label{app:ssh-odd-length}

In the main text, we give the solution of the system for even system sizes.
Here we provide the solution for chains of odd length.
In this case, the most natural perturbing elements are $\delta_l \sqrt{t_{l,1}}\sqrt{t_{l,2}}$
and $\delta_r \sqrt{t_{r,1}}\sqrt{t_{r,2}}$.
That is, the hamiltonian takes the form
\begin{equation}
	H = \begin{pmatrix}
	0& t_{l,1} & & &&\delta_r \sqrt{t_{r,1}}\sqrt{t_{r,2}}\\
	t_{r,1} &0 &t_{l,2}& &&\\
	& t_{r,2} &\ddots &t_{l,1}&&\\
	 & & t_{r,1} & \ddots&\ddots&\\
	 &&&\ddots&\ddots&t_{l,2}\\
	 \delta_l \sqrt{t_{l,1}}\sqrt{t_{l,2}}&&&&t_{r,2}&0
	\end{pmatrix} \ .
	\label{eq:ssh_hamiltonian_odd}
\end{equation}
The method to obtain the eigenvalues is the same as for the chains of even length, that is, one
determines $\lambda_{\tilde\alpha}^2$ by considering $H^2$. The functional form of $\lambda_{\tilde{\alpha}}$ is given in Eq.~\eqref{eq:eivenvalue_app}.
The equation to determine $\tilde\alpha$ is now given by
\begin{align}
\label{eq:ssh_alphaeq_odd}
&
\lambda_{\tilde\alpha}^2 \Bigl( 
\frac{\sin(\tilde{\alpha}(N+1)/2)}{\sin(\tilde{\alpha})} -
\delta_l \delta_r
\frac{\sin(\tilde{\alpha}(N-1)/2)}{\sin(\tilde{\alpha})}\Bigr)^2 =\nonumber\\
&
\delta_r^2
t_{r,1}t_{r,2}\Bigl( \frac{t_{r,1}t_{r,2}}{t_{l,1}t_{l,2}} \Bigr)^{(N-1)/2}
+ \delta_l^2 t_{l,1}t_{l,2}\Bigl( \frac{t_{l,1}t_{l,2}}{t_{r,1}t_{r,2}} \Bigr)^{(N-1)/2}\nonumber\\
&+ 2 \delta_l \delta_r \sqrt{t_{l,1}}\sqrt{t_{r,1}}\sqrt{t_{l,2}}\sqrt{t_{r,2}}
 \ .
\end{align}
In terms of $x=e^{i \tilde\alpha}$, this equation has degree $2L$ (after multiplying with $x^L$),
with solutions coming in $(x,1/x)$ pairs, both members of a pair giving rise to the same
value of $\lambda_{\tilde\alpha}^2$.
In contrast with the case $N$ even, the spectrum is not symmetric, so for each independent solution $\tilde\alpha$,
one needs to determine the actual sign of the eigenvalue.

To obtain the correct signs of the eigenvalues, we use the fact that the equation determining
$\tilde\alpha$, Eq.~\eqref{eq:ssh_alphaeq_odd}, contains $\lambda^2_{\tilde\alpha}$.
We can use this to construct a sign, that depends on the particular value of $\tilde\alpha$.
After factoring the right hand side of equation \eqref{eq:ssh_alphaeq_odd}, one finds that
\begin{align}
&\sigma_{\tilde\alpha} = \\
&\sqrt{t_{l,1}t_{r,1} + t_{l,2}t_{r,2} + 2 \cos(\tilde\alpha) \sqrt{t_{l,1}}\sqrt{t_{r,1}}\sqrt{t_{l,2}}\sqrt{t_{l,2}}} \times
\nonumber\\
&\Bigl[ 
\frac{\sin(\tilde{\alpha}(N+1)/2)}{\sin(\tilde{\alpha})} -
\delta_l \delta_r
\frac{\sin(\tilde{\alpha}(N-1)/2)}{\sin(\tilde{\alpha})}\Bigr] \times\nonumber\\
&
\frac{(\sqrt{t_{l,1}}\sqrt{t_{r,1}}\sqrt{t_{l,2}}\sqrt{t_{r,2}})^{(N-1)/2}}
{\delta_l t_{l,1}^{N/2}t_{l,2}^{N/2}+\delta_r t_{r,1}^{N/2}t_{r,2}^{N/2}}\nonumber
\end{align}
is indeed a sign, that depends on the particular value of $\tilde\alpha$ and the parameters of the model.
By studying chains of small, odd length numerically, we convinced ourselves that the sign
$\sigma_{\tilde\alpha}$ gives the correct sign of the eigenvalues, which can hence be written
as
\begin{align}
&\lambda_{\tilde\alpha} = \\
&\sigma_{\tilde\alpha}
\sqrt{t_{l,1}t_{r,1} + t_{l,2}t_{r,2} + 2 \cos(\tilde\alpha) \sqrt{t_{l,1}}\sqrt{t_{r,1}}\sqrt{t_{l,2}}\sqrt{t_{l,2}}} \ .
\nonumber
\end{align}
We note that the evaluation of $\sigma_{\tilde\alpha}$ can be somewhat unstable numerically, for given values of $\tilde\alpha$, in particular when $\delta_l$, $\delta_r$ are large in comparison to the hopping parameters in the model.

\subsection{Zero mode for even length chains}
\label{app:ssh-zero-mode-even-length}

It is well known that for open chains of {\em odd} length, there is an exact zero mode,
regardless of the parameters of the model.
This also easily follows from Eq.~\eqref{eq:ssh_alphaeq_odd} by setting $\delta_l = \delta_r = 0$.
For open chains of {\em even} lengths, there can be a zero mode, with an energy that
is exponentially small in system size.
Here, we derive the region in parameter space, for which this zero mode occurs.
We do this in a different way compared to \cite{KuEdBuBe2018,YaJiBe2022}.
Namely, we find the criterion for $\tilde\alpha$, in order to have an exact zero eigenvalue, and
determine for which parameters the equation for $\tilde\alpha$ is satisfied, up to corrections that are
exponentially small in system size.

We introduce the ratio $r_1 = \frac{\sqrt{t_{l,2}}\sqrt{t_{r,2}}}{\sqrt{t_{l,1}}\sqrt{t_{r,1}}}$, and write $x = e^{i\tilde\alpha}$
as usual.
The condition for having a zero eigenvalue then becomes $x + \frac{1}{x} = -(r_1+\frac{1}{r_1})$, with solutions
$x=-r_1$ and $x = -\frac{1}{r_1}$.
For $\delta = 0$, Eq.~\eqref{eq:ssh_alphaeq} for $\tilde\alpha$ becomes
\begin{equation}
\frac{\sin(\tilde{\alpha}(N/2+1))}{\sin(\tilde{\alpha})} +
\frac{
\sqrt{t_{l,2}}\sqrt{t_{r,2}}
}
{
\sqrt{t_{l,1}}\sqrt{t_{r,1}}
}
\frac{\sin(\tilde{\alpha}N/2)}{\sin(\tilde{\alpha})} = 0 \ ,
\end{equation}
which in terms of $x$ and $r_1$ reads (recall that $N$ is even)
\begin{equation}
x^{-N/2} \bigl( 1 + r_1 x + x^2 + r_1 x^3 + \ldots + r_1 x^{N-1} + x^{N} \bigr) = 0 \ . 
\end{equation}
For both conditions to have a zero mode, $x = -r_1$ or $x = -1/r_1$, the LHS evaluates to $r_1^{-N/2}$.
This means that the equation is satisfied in the large $N$ limit, provided that $| r_1 | > 1$.
In terms of the original parameters in the model, we find that there is a zero mode, provided that
$\left| \frac{t_{l,2} t_{r,2}}{t_{l,1} t_{r,1}} \right| > 1$.

\section{Some 2D models square lattice models}
\label{app:trival-square-lattice}

As stated in the main text in Sec.~\ref{sec:two-dimensional-systems}, it is possible to construct square lattice models by `interlacing' solvable, one-dimensional models.
These 2D models can be solved trivially, given the solutions for the one-dimensional models.
We start with a one-dimensional model given by a non-Hermitian matrix $A$ that we have solved, which depends on some hopping parameters $\{ t \}$ and perturbation $\delta$. 
By this we mean that we know both the functional form of the eigenvalues $\lambda_\alpha$ and the form of the equation that $\alpha$ should satisfy, which we denote by ${\rm eq}_\alpha$.
That is, we know there exists a matrix $V$, such that $D = V^{-1} A V$, where $D$ is diagonal, with diagonal elements
$(\lambda_{\alpha_1},\lambda_{\alpha_2},\ldots,\lambda_{\alpha_{N_1}})$ (for simplicity, we assume the system is diagonalizable).
We can now stack $N_2$ such models, and form a square lattice, by coupling them, by (say) forming HN-models in the second direction.
This gives rise to the following two-dimensional model
\begin{equation}
    H = \begin{pmatrix}
    A & d I & 0 & \cdots & 0 & \delta_2 u I\\
    u I & A & d I \\
    0 & u I & A & d I \\
    \vdots & & &\ddots\\
    0 & & & u I & A & d I \\
    \delta_2 d I & & & & u I & A
    \end{pmatrix}
\end{equation}
We can diagonalize the $A$ blocks, by means of the transformation $H' = W^{-1} H W$, where $W$ is given by $W = I_{N_2} \otimes V$.
This results in
\begin{equation}
    H' = \begin{pmatrix}
    D & d I & 0 & \cdots & 0 & \delta_2 u I\\
    u I & D & d I \\
    0 & u I & D & d I \\
    \vdots & & &\ddots\\
    0 & & & u I & D & d I \\
    \delta_2 d I & & & & u I & D
    \end{pmatrix} \ .
\end{equation}
We now observe that $H'$ is similar to the matrix
\begin{equation}
    H'' = \begin{pmatrix}
    H_1 \\
    & H_2 \\
    && \ddots \\
    &&& H_{N_1}
    \end{pmatrix} \ ,
\end{equation}
where the $H_i$ take the form
\begin{equation}
    H_i = \begin{pmatrix}
    \lambda_{\alpha_i} & d & 0 & \cdots & 0 & \delta_2 u \\
    u & \lambda_{\alpha_i} & d \\
    0 & u & \lambda_{\alpha_i} & d \\
    \vdots &&& \ddots \\
    0 &&& u &\lambda_{\alpha_i} & d\\
    \delta_2 d & &&& u & \lambda_{\alpha_i}
    \end{pmatrix} \ .
\end{equation}
As expected, this takes the form of a HN-model, with known eigenvalues $\lambda_\beta$ and known equation for $\beta$, which we here denote by ${\rm eq}_\beta$.

So, in general, the eigenvalues of this model are given by $\lambda_{\alpha,\beta} = \lambda_{\alpha} + \lambda_{\beta}$, where $\alpha$ and $\beta$ run over the independent solutions of ${\rm eq}_\alpha$ and ${\rm eq}_\beta$.
Here, we showed this result for a square lattice, with an arbitrary model we can solve in the one direction, and the HN-model in the other direction.
The result straightforwardly generalizes to a square lattice, with two arbitrary, solved models, one in each direction.
The reason why this works is trivial, simply because for the current setup, the two directions are completely independent of one another.
Clearly, this construction can be done in arbitrary dimension.

\end{document}